\documentclass[USenglish,twocolumn]{article}

\usepackage[utf8]{inputenc}
\usepackage[big]{dgruyter}
 
\usepackage{cite}
\usepackage{amsmath,amssymb,amsfonts}
\usepackage{textcomp}
\usepackage{url}
\usepackage{subfig}
\usepackage{multirow}
\usepackage{algorithm}
\usepackage{algpseudocode}
\usepackage{mathtools}
\usepackage{color}
\usepackage{multirow}

\newcommand{\OUEthicalRule}{The collection experiment of data on the in-home activities of users and sensor values in real homes received approval from the Research Ethics Committee of the Graduate School of Information Science and Technology, Osaka University.}
\newcommand{\subfigwidth}{0.25\linewidth}

\begin{document}

    \articletype{Research Article}

    \author*[1]{Masaaki Yamauchi}
    \author[2]{Masahiro Tanaka}
    \author[2]{Yuichi Ohsita}
    \author[2]{Masayuki Murata}
    \author[3]{Kensuke Ueda}
    \author[4]{Yoshiaki Kato}

    \affil[1]{Graduate School of Information Science and Technology, Osaka University - Suita, Osaka, 565-0871 Japan;
    e-mail: m-yamauchi@ist.osaka-u.ac.jp.}
    \affil[2]{Graduate School of Information Science and Technology, Osaka University - Suita, Osaka, 565-0871 Japan;
    e-mail: \{y-ohsita, murata\}@ist.osaka-u.ac.jp.}
    \affil[3]{Advanced Technology R\&D Center, Mitsubishi Electric Corporation - Amagasaki, Hyogo, 661-8661, Japan;
    e-mail: Ueda.Kensuke@ce.MitsubishiElectric.co.jp.}
    \affil[4]{Information Technology R\&D Center, Mitsubishi Electric Corporation - Kamakura, Kanagawa, 247-8501 Japan;
    email: Kato.Yoshiaki@dh.MitsubishiElectric.co.jp.}

    \title{\huge Smart-home anomaly detection using combination of in-home situation and user behavior}

    \runningtitle{Anomaly detection using in-home situation and behavior}
    \runningauthor{M. Yamauchi et al.}


    \begin{abstract}{ 
Internet-of-things (IoT) devices are vulnerable to malicious operations by attackers, which can cause physical and economic harm to users; therefore, we previously proposed a sequence-based method that modeled user behavior as sequences of in-home events and a base home state to detect anomalous operations.
However, that method modeled users' home states based on the time of day; hence, attackers could exploit the system to maximize attack opportunities.
Therefore, we then proposed an estimation-based detection method that estimated the home state using not only the time of day but also the observable values of home IoT sensors and devices.
However, it ignored short-term operational behaviors.
Consequently, in the present work, we propose a behavior-modeling method that combines home state estimation and event sequences of IoT devices within the home to enable a detailed understanding of long- and short-term user behavior.
We compared the proposed model to our previous methods using data collected from real homes.
Compared with the estimation-based method, the proposed method achieved a 15.4\% higher detection ratio with fewer than 10\% misdetections.
Compared with the sequence-based method, the proposed method achieved a 46.0\% higher detection ratio with fewer than 10\% misdetections.
    }\end{abstract}



    \journalname{Open Comput. Sci.}

\DOI{DOI}
	\startpage{1}
	\received{Sept 29, 2021}
	\revised{..}
	\accepted{..}

	\journalyear{yyyy}
	\journalvolume{vv}
	\journalissue{nn}
 
\maketitle


\section{Introduction}
Smart homes with multiple internet-connected home appliances have become widespread as part of the internet of things (IoT).
More than 12 billion IoT devices were deployed in 2020, and it is estimated that the number of IoT appliances now surpasses the number of non-IoT versions~\cite{IoTincrease2020}.
Users can connect to their IoT devices (e.g., washing machines, home sensors, and cooking stoves) via smartphones and smartwatches.
The growth of this trend is expected to continue indefinitely~\cite{IoTincrease2020}.

However, with this growth, the risk of cyberattacks targeting home IoT devices increases~\cite{AttackIoT}.
A major type of cyberattack on home IoT devices is the distributed denial-of-service attack, which affects multiple IoT devices simultaneously based on the devices' inherent vulnerabilities~\cite{IoTPOT, HomeIoTReflectiveDDosAttack}.
Fortunately, countermeasures exist~\cite{martin2017fending, HNWattack, SDN-Firewall}.

Notably, it is very difficult to maintain the boundary security of IoT devices~\cite{IoTSecurityCost} because they employ many different communication protocols and connect to many different platforms.
Moreover, proper boundary security would be exceedingly expensive~\cite{UKReportIoTCost}.
Therefore, anomaly detection systems that comprehensively monitor a smart home or a smart factory to detect abnormal (out-of-the-ordinary) IoT behaviors (e.g., signals, operating status, and error reporting)~\cite{IoTAnalysisDetection, LearningIoTPacketsBehavior1, LearningIoTPacketsBehavior2} are needed.
For example, Sivanathan~et~al. proposed a monitoring system that analyzed legitimate behaviors of IoT devices by classifying their traffic flows~\cite{IoTAnalysisDetection}.
Distributed denial-of-service attacks on smart homes have been detected by comparing suspicious traffic with usual behaviors based on home occupancy~\cite{LearningIoTPacketsBehavior1, LearningIoTPacketsBehavior2}.

Notably, cyberattacks on IoT devices create significant additional human risks~\cite{HNWthreat}.
In particular, attacks that take control of home IoT devices are considered dangerous not only in cyberspace but also in the physical world.
For example, simultaneous attacks on high-power IoT devices can suddenly increase energy demands and lead to power outages~\cite{MadIoT}.
As a discrete example, it has also been shown that in-home IoT televisions can be hijacked from the internet~\cite{IntrudeTV}; similar attacks have been shown to affect smart phones and smart watches~\cite{IntrudeSmartWatch}.

To address attacks on home IoT devices leading to anomalous operations, we previously proposed a detection method~\cite{m-yamauchi20IEEE_TCE-IoTAnomalyDetection} that modeled the behavior of users from sequences of events in their homes to assess normal behaviors.
This sequence-based method trained its model by storing event sequences based on the time of day so that deviations from operations could be detected.
However, this sequence-based method was too simplistic, and the home state was not studied in detail; hence, it was noted that an attacker could optimize attacks by studying the time-of-day behaviors.

Subsequently, we proposed another anomaly detection method~\cite{m-yamauchi20ITC-HomeEstimationDetection} that modeled home states by estimating the sensed values and operating statuses of IoT devices.
That estimation-based method calculated the operating probability of the IoT device and assessed anomalies based on a baseline threshold.
The estimation-based method achieved a better detection accuracy than a method to detect anomalous operation based on only the time of day information.
As the estimation was based on the current home situation, it was difficult for attackers to exploit the system because they could not easily estimate the timing when an attack would be likely to succeed.
However, this estimation-based method could not grasp user activities in detail over short periods.

Therefore, in this study, we propose a detection method that models user behavior by combining state estimation and behavior sequences of in-home activities performed over short periods.
Hence, our sequence-based method can grasp the short-term activities of users in detail, whereas the estimation-based method grasps the long-term transitions of the home state.
The proposed method stores the sequences in the estimated home states.
Then, the proposed method calculates the occurrence probabilities of sequences, including detection target operations, and it detects anomalous operations when the probability is lower than a threshold value.

We simulated the proposed method and compared the results to those of the previous sequence-based and estimation-based methods using datasets of behaviors and sensor values collected from real homes.

The remainder of this article is organized as follows.
We describe anomaly detection methods for operations of home IoT devices in Section~\ref{sec:related_work}.
The proposed method, including the estimation of the in-home situation, storage of behavior sequences, and their combinations, is described in Section~\ref{sec:method}.
Then, we report on the evaluation of the proposed method and the corresponding results in Section~\ref{sec:Evaluation}.
Finally, we conclude the paper and discuss possible avenues for future research in Section~\ref{sec:Conclusion}.


\section{Related Works}\label{sec:related_work}
Here, we explain detection methods of anomalous operations that learn user behaviors based on their usage of home IoT devices.

Ramapatruni~et~al. proposed a method to detect anomalous operations.
Their method used hidden Markov modeling (HMM) to learn a single user's normal activities.
HMM parameters were then trained with information obtained from IoT sensors.
Then, the trained HMM detected anomalous operations when the probability of that operation occurring was lower than a baseline threshold.
The accuracy of this method was demonstrated using a dataset collected from a smart-home environment.
The authors collected detailed activity information on the user entering and leaving the home and the operations of the consumer electronics therein.
Additionally, IoT activities from the living room, bedroom, bathroom, and closet devices were recorded.
This method learned the behaviors of a single user in detail. However, the method could not be applied to a home containing multiple users~\cite{compared_hmm}; it was examined in our previous work~\cite{m-yamauchi20IEEE_TCE-IoTAnomalyDetection}.
It is difficult to deploy this method in real homes because most involve multiple users, which greatly increases the difficulty.
In contrast, our proposed method models the situation while focusing on the states of the home instead of the states of the user.
Furthermore, the proposed method uses information that can be easily collected from commercially available IoT sensors and home gateways.
Therefore, our proposed method can be applied to real home environments.

We previously proposed a method to detect anomalous operations even in cases of multiple users by utilizing their sequence of behaviors~\cite{m-yamauchi20IEEE_TCE-IoTAnomalyDetection}.
This sequence-based method detected anomalous operations at the home gateway, which was connected to all home IoT devices, sensors, and smartphones.
The home gateway collected two types of information: the state information of the operations of devices (e.g., time of day, room temperature, and humidity) and the presence or absence of users in the home based on the statuses of their smartphones.
The home gateway subsequently classified the states of the home by constructing a table of sensed values, storing the sequences of operations of IoT devices and data on the entry and exit of users in each cell.
Finally, the home gateway judged whether legitimate or anomalous operations occurred by comparing sequences of current operations to the stored sequences of the current state.
This sequence-based method handled cases of multiple users by constructing their sequences from the monitored operations.
However, the sequence-based method used only the time-of-day information in the table to classify the states.
Therefore, an attacker could estimate the optimal attack times based on the time of day. Owing to the large impact of sequence information utilization, the sequence-based method achieved high accuracy.
However, the detailed analysis of state learning was deficient.

Hence, we proposed another anomaly detection method that estimated the states of a home based on the sensed values and operating statuses of IoT devices.
This estimation-based method calculated the operating probability of each state and detected anomalies when the probability was low.
The study compared this estimation-based method to one that used only time-of-day information, confirming that the estimation-based method was more accurate~\cite{m-yamauchi20ITC-HomeEstimationDetection}.
Therefore, an attacker could not exploit the system based only on time-of-day information.
However, the method could not learn the short-term behavior patterns of users.
Additionally, some details needed to be corrected, owing to the inadequate observed data.

In this study, we propose a more accurate detection method that combines the estimation- and sequence-based methods.
The proposed method determines the current state in the home via estimation and the state transitions by learning the behavior patterns of users. Furthermore, we improve the estimation-based method~\cite{m-yamauchi20ITC-HomeEstimationDetection} to achieve higher accuracy by fixing details and correcting the observed data.


\section{Anomaly Detection Method based on In-home Situation and Behavior Sequence}\label{sec:method}
We propose a new model that learns the behaviors of IoT users in a home to detect anomalous device operations as a safeguard against cyberattacks.
The model learns sequences of user behaviors alongside corresponding states of the home.
When an operation does not match the trained model, the model flags the operation as anomalous.

\subsection{Models used for detection}\label{sec:model}
\begin{figure}[ht]
	\centering
	\includegraphics[width=1.\columnwidth]{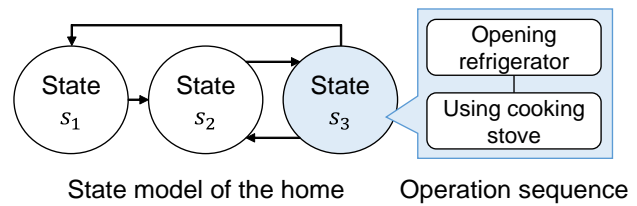}
	\caption{Overview of the training model of the proposed method. According to the estimated home state ``State $s_3$'', a cooking stove is used after opening a refrigerator.\label{fig:model_overview}}
\end{figure}

The proposed method estimates home states and stores behavior sequences collected over time.
First, it defines a timeslot scheme and updates the home status in each slot throughout the day.
During training, the model calculates the state transition probability, $a$, and the operation probability, $b$, using the labeled home state as the training data.
The method then estimates the home state by calculating the state probability of the training data using $a$ and $b$.
Following the calculation, the operation sequences of the home IoT devices are stored according to the estimated home state.
High-probability sequences are considered legitimate behaviors.
The overview of the learning model is described in Fig.~\ref{fig:model_overview}.
After storing the sequence, the proposed method calculates the probability $b'$ of each behavior sequence that occurs in real time.

Next, we describe the components of the proposed model.

\subsubsection{State of the home}\label{sec:def_states}
The proposed method labels and estimates the current home state, $s_{u,d}$, which is combined with the activity states of the users, $u$, and the usage states of the devices, $d$, obtained using the sensor values of IoT operations.

The activity state of users, $u$, reflects the situation of users in the home (e.g., all users are away, at least one user is home and active, or all users are sleeping).
Variable $u$ is thus defined several ways according to the home environment and the number and attributes of users.
For example, ``active,'' ``out,'' ``sleeping,'' etc. can be considered; thus, we can set $u$ as $u\in\{\textit{active},\textit{out},\textit{sleep}, \dots\}$.

There are four types of usage states, $d$, related to how home IoT devices are used. The state prior to use is $\textit{before}$, the state after use is $\textit{after}$, the state during use is $\textit{use}$, and others are $\textit{none}$.
Thus, the device state, $d$, is defined as $d\in \{\textit{use}, \textit{before}, \textit{after}, \textit{none}\}$.

Furthermore, the proposed method calculates the state transition probability $a$ to forecast the changes in the home state over time.
Because user behaviors differ greatly during day and night, the state transition probability, $a$, in the home varies depending on the time of day.
Therefore, in our model, the state transition probabilities are defined for each timeslot of the day.
The transition probability $a_k(i,j)$ from state $i$ to state $j$ in the $k$-th timeslot of each day is defined by Equation~(\ref{equ:def_a}), where $s_k$ is the state in the $k$-th timeslot.

\begin{equation}\label{equ:def_a}
	a_k(i, j) = P( s_{k} = j | s_{k-1} = i ).
\end{equation}

Additionally, the proposed method calculates the operation probability, $b$, to reflect the activity of the users to the home states as devices are operated.
The operation probability~$b$ differs for each state~$i$ and for each operation~$x$.
Therefore, in this model, the operation probabilities are defined for each state.
The operation probability~$b(i,x)$ of the operation~$x$ in state~$i$ is defined by the following Equation~(\ref{equ:def_b}):
\begin{equation}\label{equ:def_b}
	b (i, x) = P ( x | s = i ). 
\end{equation}

\subsubsection{Sequence of events in the home}
The event sequences in the home are stored and used for detection.
The information is obtained from the home network via IoT operation packets, including information of the connection and disconnection of smartphones.

As with the sequence-based method, an event sequence is defined as a series of events performed within $T_{\textit{seq}}$~s, where $T_{\textit{seq}}$ is a parameter determining whether or not events are considered as a sequence.
We consider actions A and B to be a series if they satisfy Equation~(\ref{sec:def_T_seq}), where $\textit{time}_A$ and $\textit{time}_B$ are the times when actions A and B are performed and $\textit{Diff}(\textit{time}_A, \textit{time}_B)$ is a function that obtains differences in seconds between $\textit{time}_A$ and $\textit{time}_B$.

\begin{equation}\label{sec:def_T_seq}
    \textit{Diff}(\textit{time}_A, \textit{time}_B) \leq T_{\textit{seq}}.
\end{equation}

Furthermore, after storing the sequences in the estimated states, the proposed method calculates the behavior sequence probability~$b'$ for detection.
When the proposed method identifies an event sequence including the operations of the detection target device, the probability of occurrence of sequences is calculated by multiplying the state probabilities by the behavior sequence probability~$b'$.
The behavior sequence probability~$b'(i,y)$ in state~$i$ of the sequences~$y$ is defined by Equation~(\ref{equ:def_b_dash}).

\begin{equation} \label{equ:def_b_dash}
    b'(i, y) = P ( y | s' = i ),
\end{equation}

where $s' = i$ means that the estimated state is $i$.


\subsection{Training the model}\label{sec:training}
The proposed method trains the model using data collected from the home divided into timeslots.
We assign the observed values and labels of the home states~$s_{u,d}$ to the timeslots.
Then, the proposed method calculates the state transition probability, $a_k(i,j)$, according to the labeled states, $s_{u,d}$.
The proposed method also calculates the operation probability~$b(i,x)$ that the device is operated in each state.
Next, the proposed method calculates the state probability~$\alpha(s)$ with $a$ and $b$.
The proposed method generates multiple sequences assuming that the homes have multiple users.
Based on the calculated state probability~$\alpha(s)$ and the generated sequences, event sequences are stored for each estimated home state.
Then, the proposed method calculates the operation probability~$b'(i,y)$ in each estimated state.
The rest of this subsection explains the details.

\subsubsection{Labeling the training dataset}\label{sec:labeling_method}
To create the training data, we divided the observed data into multiple parts using timeslots and labeled the states accordingly.
The state~$s$ is set by combining the activity state of the users~$u$ and the usage state of the device~$d$ as defined.

The state of the users~$u$ is set using IoT sensor data according to predefined rules.
Because these rules vary depending on the target device, the type of IoT sensors, and the number of users, we set the labeling rule according to the scenario.

\begin{figure}[ht]
	\centering
	\includegraphics[width=1.\columnwidth]{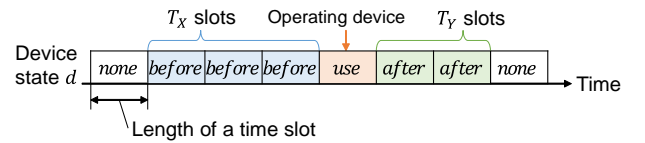}
	\caption{Labeling rule of device state~$d$, where $T_X$ is $3$ and $T_Y$ is $2$.}
	\label{fig:device_state}
\end{figure}

The device state~$d$ is determined based on the time the target device is operated.
As shown in Fig.~\ref{fig:device_state}, we define the four states of the target device~$d$ as follows.
$\textit{use}$ indicates that the device is in use, and $\textit{before}$ indicates that the device will be used within $T_X$ timeslots.
Similarly, $\textit{after}$ indicates that the device has been used within $T_Y$ timeslots, while $\textit{none}$ denotes other states.
Variables $T_X$ and $T_Y$ are parameters.

Furthermore, to change the state for each device operation, we update them as observed. An example of the learning data is described in Table~\ref{tab:log_sample}.

\begin{table*}[htb]
    \centering
    \caption{Learning data samples include the observed information, the labeled states, and variable characters. The labels change even in the same timeslot according to the operations of the devices. In this table, we set the length of the timeslot to 1~min. As a sample rule for the state of users~$u$, we set $\textit{sleep}$ when the CO2 value is higher than $35$ and the noise value is lower than $1,500$. As a sample rule of the device state~$d$, we set $T_X$ as $1$ and $T_Y$ as $1$.}
    \label{tab:log_sample}
    \begin{tabular}{l|cc|ccc|ccc}
        &\multicolumn{2}{c|}{Date information} & \multicolumn{3}{c|}{Observed information} & \multicolumn{3}{c}{Labeled states} \\
        \hline
        \multirow{2}{*}{ID}&\multirow{2}{*}{Date} & $k$-th timeslot of the day & \multirow{2}{*}{CO2} & \multirow{2}{*}{Noise} & \multirow{2}{*}{Operation} & Users & Device & Home \\
        & & /$t$-th timeslot of the data &  &  &  &  $u$ & $d$ & $s_{u,d}$ \\
        \hline
        4350&2020/1/3 23:56:00 & 1438/4318 & 34 & 1520 & --- & $\textit{active}$ & $\textit{none}$ & $s_{\textit{active},\textit{none}}$ \\
        4351&2020/1/3 23:57:00 & 1439/4319 & 34 & 1520 & --- & $\textit{active}$ & $\textit{before}$ & $s_{\textit{active},\textit{before}}$ \\
        4352&2020/1/3 23:58:00 & 1440/4320 & 34 & 1520 & --- & $\textit{active}$ & $\textit{before}$ & $s_{\textit{active},\textit{before}}$ \\
        4353&2020/1/3 23:58:20 & 1440/4320 & 34 & 1520 & Refrigerator\_Open & $\textit{active}$ & $\textit{before}$ & $s_{\textit{active},\textit{before}}$ \\
        4354&2020/1/3 23:58:35 & 1440/4320 & 34 & 1520 & Cooking oven\_On & $\textit{active}$ & $\textit{using}$ & $s_{\textit{active},\textit{cooking}}$ \\
        4355&2020/1/3 23:59:00 & 1/4321    & 34 & 1520 & --- & $\textit{active}$ & $\textit{after}$ & $s_{\textit{active},\textit{after}}$ \\
        4356&2020/1/4 00:00:00 & 2/4322    & 41 & 1480 & --- & $\textit{sleep}$ & $\textit{none}$ & $s_{\textit{sleep},\textit{none}}$ \\
    \end{tabular}
\end{table*}

\subsubsection{Calculating state transition probability and the operation probability} \label{sec:calculate_a_and_b}
Based on the labeled home states for changing timeslots, we calculate state transition probabilities~$a_k(i,j)$ from state~$i$ to state~$j$ at the $k$-th timeslot during the day.
This is used to calculate the probability~$\alpha_t(s)$ that the home state~$s$ is in timeslot~$t$.
Although the time of the state transition varies daily, similar state transitions occur in similar timeslots.
Therefore, $a_k(i,j)$ is calculated by Equation~(\ref{equ:a}) by considering the data from the $k-T_Z$-th timeslot to the $k+T_Z$-th timeslot of each day.

\begin{equation} \label{equ:a}
	a_k(i, j) =
    \begin{dcases}
        \dfrac
        { \displaystyle \sum_{m=k-T_Z}^{k+T_Z} N_{m+1,j} }
        { \displaystyle \sum_{m=k-T_Z}^{k+T_Z} N_{m,i}   } & \left(\sum_{m=k-T_Z}^{k+T_Z} N_{m,i} \neq 0\right) \\
        0   & \left(\sum_{m=k-T_Z}^{k+T_Z} N_{m,i} = 0\right).
    \end{dcases}
\end{equation}

The variable~$N_{m,i}$ represents the number of timeslots in the training data at the $m$-th timeslot of the day, the state of which is labeled as~$i$.
$T_Z$ denotes the number of similar timeslots around the target.
Parameter~$T_Z$ has a different value for each $k$; thus, we set the minimum value that satisfies $\sum_{m=k-T_Z}^{k+T_Z} N_{m,i}\neq 0$ for all states~$i$.
Even if $\sum_{m=k-T_Z^{\textit{max}}}^{k+T_Z^{\textit{max}}} N_{m,i}\neq 0$ is not satisfied for all states~$i$, where $T_Z^{\textit{max}}$ is the maximum value for $T_Z$, we set $T_Z$ as $T_Z^{\textit{max}}$.

We next explain how to calculate the operation probability~$b$ which is used to correct the state probability~$\alpha$.
Operation probability~$b(i,x)$ denotes the probability of the number of the operations~$x$ of the IoT device in state~$i$.
$b(i,x)$ is calculated using Equation~(\ref{equ:b}):

\begin{equation}\label{equ:b}
	b (i, x) = 
    \begin{dcases}
        \frac{ \sum_{k} \displaystyle N_{k,i}^{(x)} }{ \displaystyle \sum_{k} N_{k,i} } & \sum_{k} N_{k,i} \neq 0 \\
        0 & \sum_{k} N_{k,i} = 0.
    \end{dcases}
\end{equation}

Note that $N_{k,i}^{(x)}$ represents the number of occurrences of operation~$x$ in the state~$i$ in the $k$~th timeslot of the day.
If there are no operations~$x$ in the training data, $b(i,x)$ is set to $1$ for all states~$i$ to avoid incorrect transitions.

\subsubsection{Calculating state probability}\label{sec:calculate_alpha}
The proposed method calculates the state probability~$\alpha$ for each timeslot of the training data by the calculated $a$ and $b$.
Then, the proposed method stores the sequences of home events using the training data by the estimated home states because the proposed method stores sequences not only in the current home state but also in similar states.
To determine the similar states, we use the calculated state probability.
By storing sequences in the states that satisfy the conditional probability expression, we can store sequences in the similar states.

When the timeslot changes, the state transitions from the state of the previous timeslot using the learned state transition probability, $a$.
First, the proposed method calculates $\hat{\alpha}_t(i)$, the probability of state~$i$ when the timeslot is changed to~$t$, using the learned state transition probability~$a_k(i,j)$.

\begin{equation}\label{equ:alpha_hat_a}
	\hat{\alpha}_{T(t)}(i) = \sum_{j} a_{K(t)}(j,i) \alpha_{(T(t)-\Delta T)}(j).
\end{equation}

Variable $K(t)$ is a function that returns the corresponding $K(t)$-th timeslot of the day with timeslot $t$; $T(t)$ is a function that returns the time corresponding to timeslot~$t$, and $\Delta T$ indicates a very small time.
By considering the case that, in the previous timeslot, the state probability, $\alpha$, is updated by using the operation probability, $b$, Equation~(\ref{equ:alpha_hat_a}) uses the state probability at $T(t)-\Delta T$.
Then, the proposed method calculates $\alpha$ based on $\hat{\alpha}$ so that the sum of the state probabilities of each state is $1$ using Equation~(\ref{equ:set_sum_to_1}):

\begin{equation}\label{equ:set_sum_to_1}
	\alpha_{T(t)}(i) = \frac { \hat{\alpha}_{T(t)}(i) } { \sum_{j} \hat{\alpha}_{T(t)}(j) }.
\end{equation}

When we observe an operation~$x$ of a home IoT device, the proposed method updates the state probability~$\alpha$ using the operation probability~$b(i,x)$.
First, the proposed method calculates~$\hat{\alpha}_{T(x)}(i)$ according to Equation~(\ref{equ:alpha_hat_b}).

\begin{equation}\label{equ:alpha_hat_b}
    \hat{\alpha}_{T(x)}(i) = b(i,x) \alpha_{(T(x)-\Delta T)}(i),
\end{equation}

where $T(x)$ represents the time when the proposed method observed operation~$x$.
Then, the proposed method calculates the state probability, $\alpha$, after the operation of the home IoT device using Equation~(\ref{equ:set_sum_to_1}).

\subsubsection{Storing sequences}\label{sec:storing_sequence}
Based on the calculated state probability~$\alpha$, the proposed method stores the behavior sequences to estimated states. First, we must generate the sequences based on the observed operations and the users entering and leaving.
This will account for multiple users operating devices within $T_{\textit{seq}}$~s of each other.
When the users operate devices from their respective smartphones, we can identify correct behavior sequences by classifying those who operated which home IoT device based on the IP address of the operating smartphones. There are many cases where it is impossible to distinguish which user performs each operation.
Thus, as with the sequence-based method~\cite{m-yamauchi20IEEE_TCE-IoTAnomalyDetection}, we generate multiple types of sequences from a simple series of events by removing some of them for training.
For example, when actions A, B, and C are performed within $T_{\textit{seq}}$~s, equations $\textit{Diff}(\textit{time}_A,\textit{time}_B)\leq T_{\textit{seq}}$, $\textit{Diff}(\textit{time}_A,\textit{time}_C)\leq T_{\textit{seq}}$, and $\textit{Diff}(\textit{time}_B,\textit{time}_C)\leq T_{\textit{seq}}$ are satisfied.
$\textit{time}_A$, $\textit{time}_B$, and $\textit{time}_C$ represent the times when actions A, B, and C are performed, respectively.
In this example case, we generate and use all seven types of event sequences: A-only, B-only, C-only, A-B, B-C, A-C, and A-B-C.
If actions A and B are performed by the same user, and action C is performed by another, the correct event sequences, A-B and C-only, are learned.
However, incorrect sequences, such as A-only, B-only, A-C, B-C, and A-B-C are also stored.
If sequences A-B and C-only are frequently performed by users, the correct sequences will be stored multiple times.
Therefore, by using only the sequences that are greater than or equal to a given threshold, we can identify frequent behaviors.

After generating the sequences, event sequence~$y$, which is related to the operation of the detection target home IoT device, is stored for each state in which the sequences are performed.
We can determine the states for which the proposed method stores the sequences from the calculated probability, $\alpha_t(i)$, in state~$i$ at timeslot~$t$.
We select either Equation ~(\ref{equ:storing_sequence_threshold}) or~(\ref{equ:storing_sequence_rank}) and store the sequences into all states satisfying the selected one.

\begin{equation}\label{equ:storing_sequence_threshold}
	\alpha_{(T(y)-\Delta T)}(i) \leq L_{\alpha}, \\
\end{equation}

\begin{equation}\label{equ:storing_sequence_rank}
    \textit{Rank}(\alpha_{(T(y)-\Delta T)}(i)) \leq L_{\textit{Rank}}. \\
\end{equation}

Note that $T(y)$ represents the time during which sequence~$y$ occurs.
Here, $\textit{Rank}(\alpha_{(T(y)-\Delta T)}(i))$ is a function that returns the number from the top of the state probability of $i$ of all states, such as 1st, 2nd, etc.
$L_{\alpha}$ and $L_{\textit{Rank}}$ are the parameters.
When there are no states satisfying the selected equation, the proposed method does not store the sequence.

After storing the sequences, we calculate the behavior sequence probability~$b'(i,y)$ in estimated state~$i$ of the sequence~$y$ for detection.
We can then calculate~$b'$ using Equation~(\ref{equ:b_dash}):

\begin{equation} \label{equ:b_dash}
    b'(i, y) =
    \begin{dcases}
        \frac{ M(i, y) } { N'_{i} } & N'_{i} \neq 0\\
        0 & N'_{i} = 0,
    \end{dcases}
\end{equation}

where $M(i,y)$ is a function that returns the occurrence times of sequence~$y$ in the estimated state~$i$ from the training data, and $N'_{i}$ is a function that outputs the number of timeslots of the training data estimated as state~$i$.


\subsection{Detection using the learned model}\label{sec:detection}
After training the model, when an event sequence, $y$, includes operations of the detection target device, the proposed method calculates the state probability~$\alpha_{(T(y)-\Delta T)}(i)$ using Equations~(\ref{equ:alpha_hat_a}), (\ref{equ:set_sum_to_1}), and~(\ref{equ:alpha_hat_b}).
The proposed method calculates the probability of occurrence~$\delta_{(T(y)-\Delta T)}(y)$ of the sequence~$y$ by multiplying the state probabilities, $\alpha_{(T(y)-\Delta T)}(i)$, by the behavior sequence probability, $b'(i,y)$, as described in Equation~(\ref{equ:delta}):

\begin{equation} \label{equ:delta}
    \delta_{(T(y)-\Delta T)}(y) = \sum_{i} b'(i,y) \alpha_{(T(y)-\Delta T)}(i).
\end{equation}

When the calculated occurrence probability, $\delta$, satisfies Equation~(\ref{equ:threshold_nl}), the proposed method detects the operation as an anomalous operation.

\begin{equation} \label{equ:threshold_nl}
    \delta_{(T(y)-\Delta T)}(y) < n_{L(y)}.
\end{equation}

Function $L(y)$ returns the length of sequence~$y$; the length of the sequence reflects the number of events comprising the sequence.
$n_{L(y)}$ is a parameter of the sequence constructed by $L(y)$~events.
We set multiple thresholds for each length of the sequence because long sequences are rare.


\section{Evaluation}\label{sec:Evaluation}
To evaluate the proposed method, we simulated anomaly detection using data from two real homes. We evaluated the effectiveness of each part by comparing the detection results to the results of alternative methods.

In this evaluation, we chose the operations of a cooking stove as the detection target device.
We prepared the proposed method according to the target device and the home environments.

\subsection{Evaluation environment}\label{sec:evaluation_environment}
Here, we describe the details of the detection simulation of the proposed and compared methods.
First, we explain how the datasets were collected.
Then, we set the proposed anomaly detection method suitable for each home by defining the states of the home and the labeling rules.
Thereafter, we describe the metrics of the comparison and present the results.

\subsubsection{Data collection in real homes}\label{sec:Data_collection}
We collected data of user behaviors and observed the values of the installed home IoT sensors from two real houses, A and~B\footnote{\OUEthicalRule}.
Home~A had two users who operated devices, and home~B had one.
We used monthly data of each home as one case for the simulation, resulting in 20~cases.
We describe the case using the data of home~A as $A_1$, $A_2$, $\dots$, $A_{10}$ and home~B as $B_1$, $B_2$, $\dots$, $B_{10}$.

First, we collected the date information of events, including operations of consumer electronics and user entry/exit statuses, as shown in Table~\ref{tab:collected_operations}.
Because each home included home appliances that were not connected to the internet, we collected their information by asking users to record their device use times.
For the simulation, we assumed that each home appliance was an IoT appliance, and the recorded operation logs were used for the purposes described.
Logs were compiled as buttons were pressed on the home appliances and when users entered and left the home.
Because there were several omissions in the collected logs, we corrected them via labeling rules, as described in Section~\ref{sec:labeling_rule}.

\begin{table}[ht]
	\centering
	\caption{Collected operations and events by our experimental system deployed in real homes.}
	\label{tab:collected_operations}
	\begin{tabular}{l|l}
		Device or event 	& Action	    \\
		\hline
		User position		& Entry / Exit	\\
		Room light			& On / Off	\\
		Air conditioner		& Cooling / Heating / Turning up /\\
							& Turning down / Off	 \\
		Electric fan		& On / Off	\\
		Heater				& On / Off	\\
		Washing machine     & On	\\
		Refrigerator		& Opening	\\
		TV					& On / Off	\\
		Cooking stove		& On / Off	\\
		Microwave 			& On	\\
		Toaster oven		& On	\\
		Rice cooker			& On	\\
	\end{tabular}
\end{table}

Then, we installed IoT sensors in each home and collected the sensor values shown in Table~\ref{tab:collected_sensor_data} in 5~min intervals.

\begin{table}[ht]
	\centering
	\caption{Collected sensor data from installed IoT sensors in real homes.}
	\label{tab:collected_sensor_data}
	\begin{tabular}{c|c}
		Sensor data			& Range of sensor values	\\
		\hline
		Room temperature	& 0 - 50\textdegree C			\\
		Humidity			& 0 - 100\%				\\
		Atmosphere			& 260 - 1,260 mbar			\\
		CO2					& 0 - 5,000 ppm				\\
		Noise				& 30 - 130 dB				\\
	\end{tabular}
\end{table}


\subsubsection{Settings of anomaly detection method} \label{sec:setting_of_anomaly_detection_method}
To simulate anomaly detection for a cooking stove, we set up the state of the home and labeling rules.
For this evaluation, the timeslot was assumed to be 1~min for capturing state transitions.

\paragraph{Setting home states}\label{sec:evaluation_setting_state}
We set the usage state of the devices~$d$ based on cooking states: $d \in \{\textit{use}, \textit{before}, \textit{after}, \textit{none}\}$ because cooking stoves are frequently used during cooking.
State $\textit{use}$ refers to the cooking state, $\textit{before}$ and $\textit{after}$ indicate times before and after cooking, respectively, and $\textit{none}$ implies other states.
Note that to grasp the cooking state exactly, we also used operations of the cooking appliances other than the cooking stove to label the states~$d$.
Specifically, when the cooking stove, microwave, toaster oven, or rice cooker was operated, we set $d$ as $\textit{use}$; the details are described in Section~\ref{sec:labeling_rule}.

We set the activity state of the users as $u \in \{\textit{active}, \textit{out}, \textit{sleep}\}$.
Hence, at least one person was active, everyone in the home was out, or everyone was sleeping, respectively.

We set the home states~$s_{u,d}$ by combining $u$ and $d$.
However, states~$s_{\textit{out},\textit{use}}$ and $s_{\textit{sleep},\textit{use}}$ did not exist because users cannot cook while they are sleeping or out of the home.
Hence, we set 10~states excluding the above for detection.

\paragraph{Labeling rule}\label{sec:labeling_rule}
Using the defined states from Section~\ref{sec:evaluation_setting_state}, we labeled each timeslot of the training data.
In consideration of privacy concerns, we labeled the home states from the observed information taken from the IoT devices and sensors.
In particular, because there were several omissions in the collected logs, we corrected them based on the rules.

The activity states of the users~$u$ are labeled as follows.
\begin{itemize}
    \item[$\textit{out}$:] The timeslots that the home was empty were tabulated by counting the number of users in the home based on their entry and exit time information. However, when we observed an operation of a home IoT device, we changed the number of users in the home to $1$ and set the states of the timeslot after the time corresponding to the change. This is because the logs included some omissions of entries and exits. In this case, we excluded logs of the day from the calculation of $a$ and $b$.
    \item[$\textit{sleep}$:] The timeslots at night containing noise values were lower than a threshold, and the CO2 concentration value was higher than a threshold; the installed IoT sensors in each home sensed the values.
    We defined the thresholds by the sleeping time that we asked of the subjects, including the noise and CO2 values of the sleeping time.
    Concretely, we defined the night from 22:00 to 9:59, the noise threshold as 35~dB, and the CO2 threshold as 1,500~ppm in home~A and 400~ppm in home~B.
    When two $\textit{sleep}$ timeslots existed within 90~min, we labeled the timeslots between the two as $\textit{sleep}$, because the indicators were temporarily lowered during sleep.
    When we observed an operation in $\textit{sleep}$ states, we corrected the states to $\textit{active}$ because more than one user was awake and active. Concretely, when a user operated devices in the time frame of 22:00 to 4:59, we changed the user states to $\textit{active}$ before 5~h from the time of operation; when a user operated devices at the time from 5:00 to 9:59, we changed the user states to $\textit{active}$ after 4~h from the timeslot during which the device was operated.
    \item[$\textit{active}$:] This refers to states other than $\textit{out}$ and $\textit{sleep}$.
\end{itemize}

Then, the usage states of device~$d$ (i.e., cooking or not) in this evaluation are labeled as follows.
\begin{itemize}
    \item[$\textit{use}$:] This refers to timeslots in which a user operates a cooking appliance, including the cooking stove, microwave, toaster oven, and rice cooker.
    Because the cooking continues for a certain time, we set the $T_C$ timeslots after the operating cooking appliances as $\textit{use}$, where the $T_C$ is a parameter of cooking time.
    We did not include the refrigerator in the cooking appliances because it is used frequently even when users are not cooking. Furthermore, when there are two $\textit{use}$ states within 15~min, we labeled the timeslots between the two as $\textit{use}$.
    \item[$\textit{before}$:] This indicates the $T_X$ timeslots before $\textit{use}$.
    \item[$\textit{after}$:] This indicates the $T_Y$ timeslots after $\textit{use}$.
    \item[$\textit{none}$:] This indicates states other than the $\textit{use}$, $\textit{before}$, and $\textit{after}$.
\end{itemize}

We labeled the home states, $s_{u,d}$, by combining the labeled states of the users~$u$ and those of the devices~$d$.


\subsubsection{Metrics} \label{sec:evaluation_metrics}
We evaluated the proposed method using two metrics: detection and misdetection ratios.
For the simulation, we mixed 100~anomalous operations of the cooking stove at random times during the day.
Furthermore, we considered the actual operations of the home IoT devices originally included in the recorded log as legitimate operations.
The detection ratio and misdetection ratio was calculated using Equation~(\ref{equ:tpr}) and~(\ref{equ:fpr}).

\begin{equation} \label{equ:tpr}
    \textrm{Detection ratio} = \frac{ \textit{TP} } { \textit{TP} + \textit{FN} }
\end{equation}
Here, $\textit{TP}$ is the number of true positives of detected anomalous operations; $\textit{FN}$ is the number of false negatives; and $\textit{TP}+\textit{FN}$ equals $100 N^{\textit{days}}$, where $N^{\textit{days}}$ indicates the number of days included in the detection data.

\begin{equation} \label{equ:fpr}
    \textrm{Misdetection ratio} = \frac{ \textit{FP} } { \textit{FP} + \textit{TN}. }
\end{equation}
Here, the $\textit{FP}$ is the number of false positives that are legitimate operations the methods could not determine as legitimate; the $\textit{TN}$ is the number of true negatives.

For the evaluation, we used cross-validation.
First, we trained the models with data for one month excluding one day.
Then, we simulated the detection of the trained model using the excluded data.
By changing the excluding day and summarizing the detection results, we obtained a detection result from the monthly data.

We changed the parameter values in each combination respectively and collected the combinations of detection and misdetection ratios.
We describe the detection results as figures with the misdetection ratio on the horizontal axis and the detection ratio on the vertical axis.
Thus, we only plotted the results having the highest values on the vertical axis among the results that were less than or equal to the values on the horizontal axis.

Note that when the operation of the target device occurred, a decision was made based on the sequence that was generated up to and just before the operation.
Hence, the operations subsequent to the target operation were not used for the detection of the target operation.


\subsubsection{Compared methods}\label{sec:compared_methods}
To evaluate the effectiveness of the proposed method, we compared it to the other methods.
Thus, we demonstrated the improvements gained by combining the sequence information.
By comparing with the sequence-based method, we confirmed the effectiveness of estimation of the in-home situation.
The differences between the proposed method and the compared methods are described in Table~\ref{tab:comparative_methods}.

\begin{table}[ht]
	\centering
	\caption{Differences between the proposed and compared methods.}
	\label{tab:comparative_methods}
	\begin{tabular}{c|c|c}
		Method                      & State         & Sequence	    \\
		\hline
		Proposed         	& Estimating situation	& \checkmark	\\
		Estimation-based~\cite{m-yamauchi20ITC-HomeEstimationDetection}& Estimating situation &\\
		Sequence-based~\cite{m-yamauchi20IEEE_TCE-IoTAnomalyDetection}& Time of day&\checkmark\\
	\end{tabular}
\end{table}


\paragraph{Estimation-based anomaly detection method}
We compared our new method to the estimation-based anomaly detection method.
This method estimates the states of the home based on the sensed values and operating statuses of IoT devices.
As with the proposed method, the estimation-based method calculated the~$b$ and~$\alpha$ using Equations~(\ref{equ:a}-\ref{equ:alpha_hat_b}).
We calculated the probability that the operation was legitimate by multiplying $b$ to $\alpha$ and by summarizing them.
If the value was higher than a threshold~$\theta$ the equation was regarded as legitimate, as shown in Equation~\ref{equ:theta}.

\begin{equation}\label{equ:theta}
    \sum_{i} \alpha_{T(x_{c}-\Delta T)}(i) b(i, x_{c}) > \theta,
\end{equation}

where $x_c$ denotes the operation of the cooking stove.


\paragraph{Sequence-based anomaly detection method}
We compared our new method to the sequence-based anomaly detection method~\cite{m-yamauchi20IEEE_TCE-IoTAnomalyDetection}.
This method models the behaviors of users from sequences of events in the home at each time of day.
The sequence includes operations performed within $T_{\textit{seq}}$~s.
When this method observes a sequence related to the detection target operation, it counts the number of stored equivalent sequences that occurred during the time of day within $\alpha_{\textit{seq}}$~s of the observed sequence.
When the ratio of the counted number of all stored operations of the target device was greater than or equal to the threshold, $n_d^{\textit{seq}}$, the target operation included in the sequence was judged as legitimate.
$n_d^{\textit{seq}}$ is the parameter, and the $d$ denotes the length of the sequence.

Anomalous operations of the cooking stove must be detected immediately because such operations present higher risks to users compared to other devices such as TVs, air conditioners, etc.
Therefore, for this evaluation, the proposed and sequence-based methods could only use the sequences leading up to the target operations.
Additionally, the cooking stove was often operated as the first event of a sequence when users wished to cook.
These points differ from the evaluation of the previous sequence-based method~\cite{m-yamauchi20IEEE_TCE-IoTAnomalyDetection}.


\subsubsection{Parameter values}\label{sec:parameters}
Training and detection were performed for each combination of values set in Table~\ref{tab:parameters}.
We simulated all combinations with each value set for each parameter and evaluated the detection results.

\begin{table}[ht]
	\centering
	\caption{Values of parameters for the proposed and compared methods.}
	\label{tab:parameters}
	\begin{tabular}{c|l}
		Parameter		& Set values	\\
		\hline
		$T_X$			& $15, 30, 60, 100.$\\
		$T_Y$			& $15, 30, 60, 100.$\\
		$T_C$           & $10, 15, 20, 30, 45, 60.$\\
	    $n_l(l=1)$               & $0.0, 1.0\times 10^{-7}, 2.0\times 10^{-7}, \dots, 1.0.$\\
	    $n_l(l\geq2)$               & $0.0, 1.0\times 10^{-7}, 2.0\times 10^{-7}, \dots, 1.0.$\\
	    $T_{\textit{seq}}$  & $600$\\
	    $L_{\alpha}$        & $1, 2, \dots, 10.$\\
		$L_{\textit{Rank}}$ & $0.00, 0.05, 0.10, \dots, 1.00.$\\
		$\alpha_{\textit{seq}}$			& $0, 900, 3600, 10800, 32400, 43200.$\\
		$n_d^{\textit{seq}} (d=1) $		& $0.00, 0.02, 0.05, 0.1, 0.15, 0.20, 0.25,$ \\
	                        & $0.30, 0.35, 0.40, 0.45, 0.50, 1.00.$\\
		$n_d^{\textit{seq}} (d\geq2)$	    & $0.00, 0.02, 0.05, 0.1, 0.15, 0.20, 0.25,$ \\
	                        & $0.30, 0.35, 0.40, 0.45, 0.50, 1.00.$\\
	   $T_Z^{\textit{max}}$ & $720$\\
	\end{tabular}
\end{table}


\subsection{Evaluation results}\label{sec:evaluation_result}
\begin{figure*}[ph]
	\centering
	\subfloat[$A_1$]{\includegraphics[width=\subfigwidth]{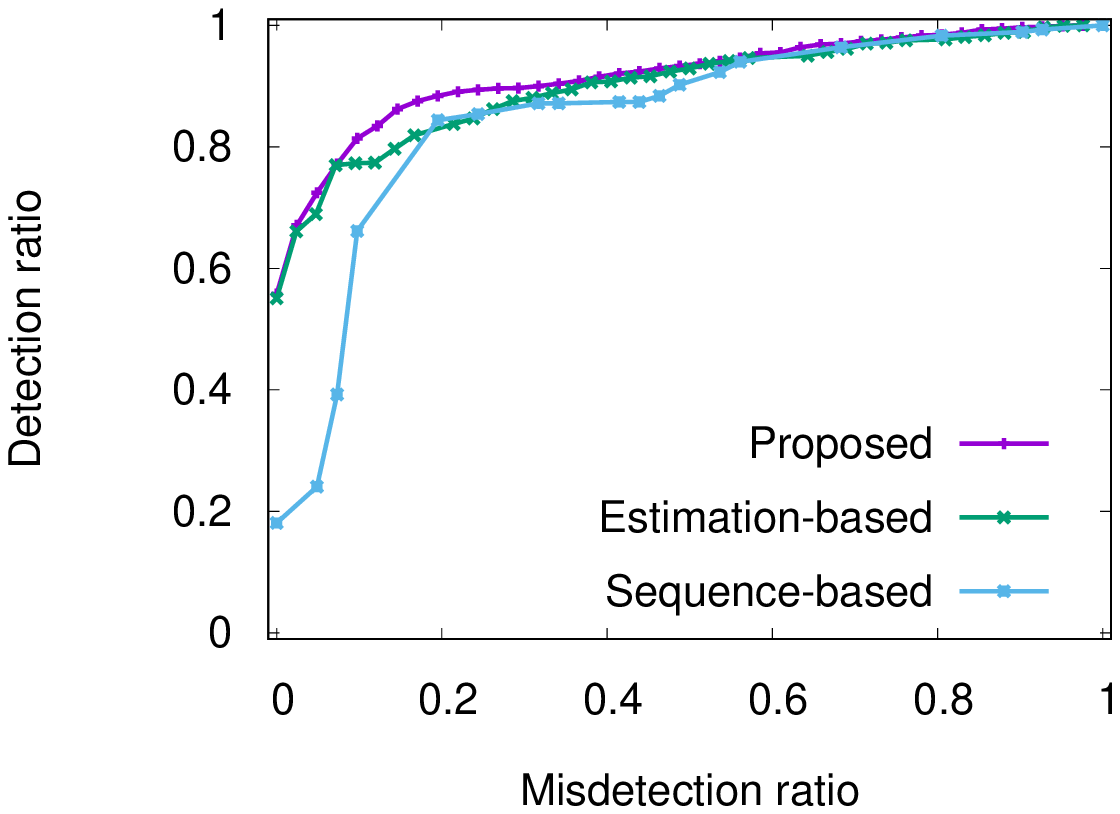}}
	\subfloat[$A_2$]{\includegraphics[width=\subfigwidth]{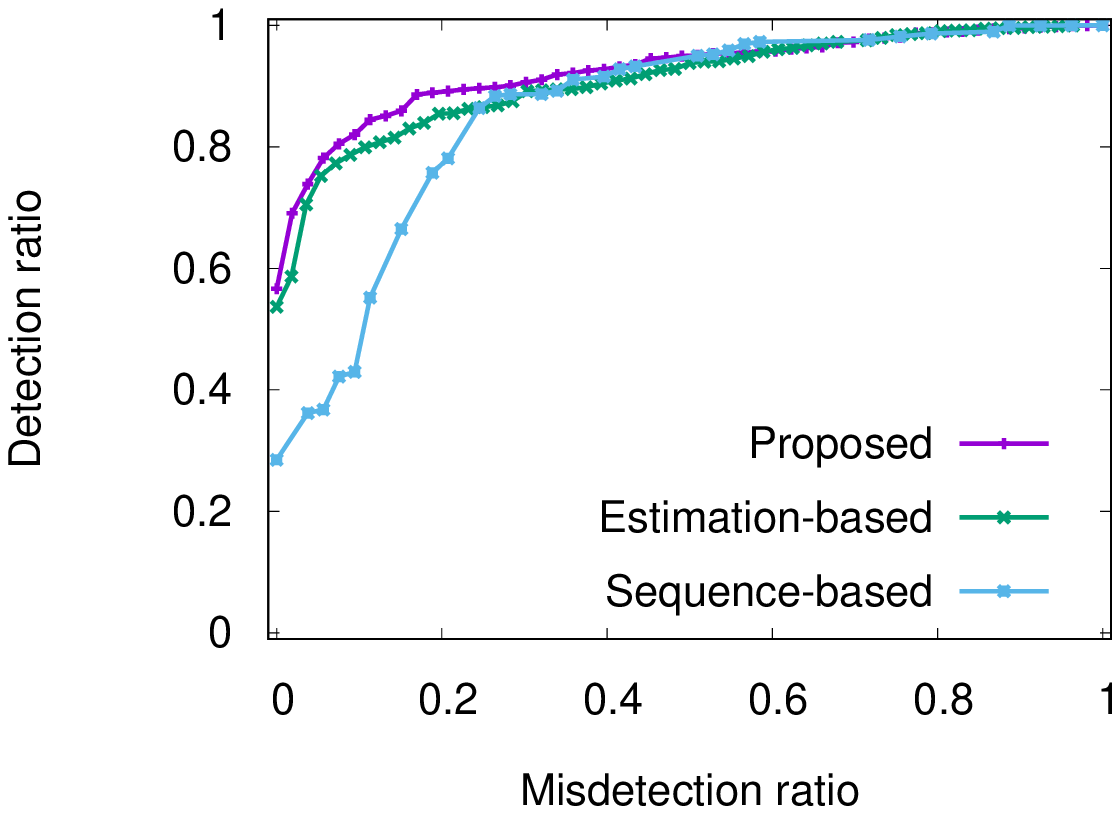}}
	\subfloat[$A_3$]{\includegraphics[width=\subfigwidth]{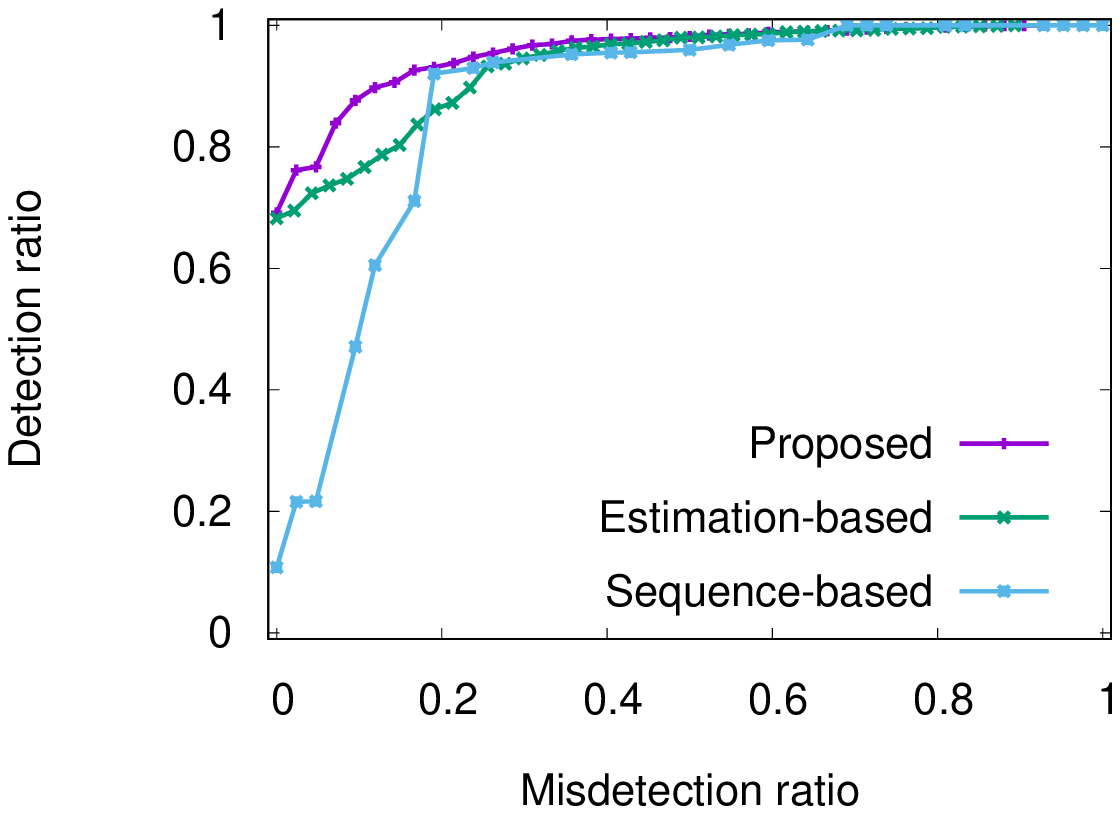}}
	\subfloat[$A_4$]{\includegraphics[width=\subfigwidth]{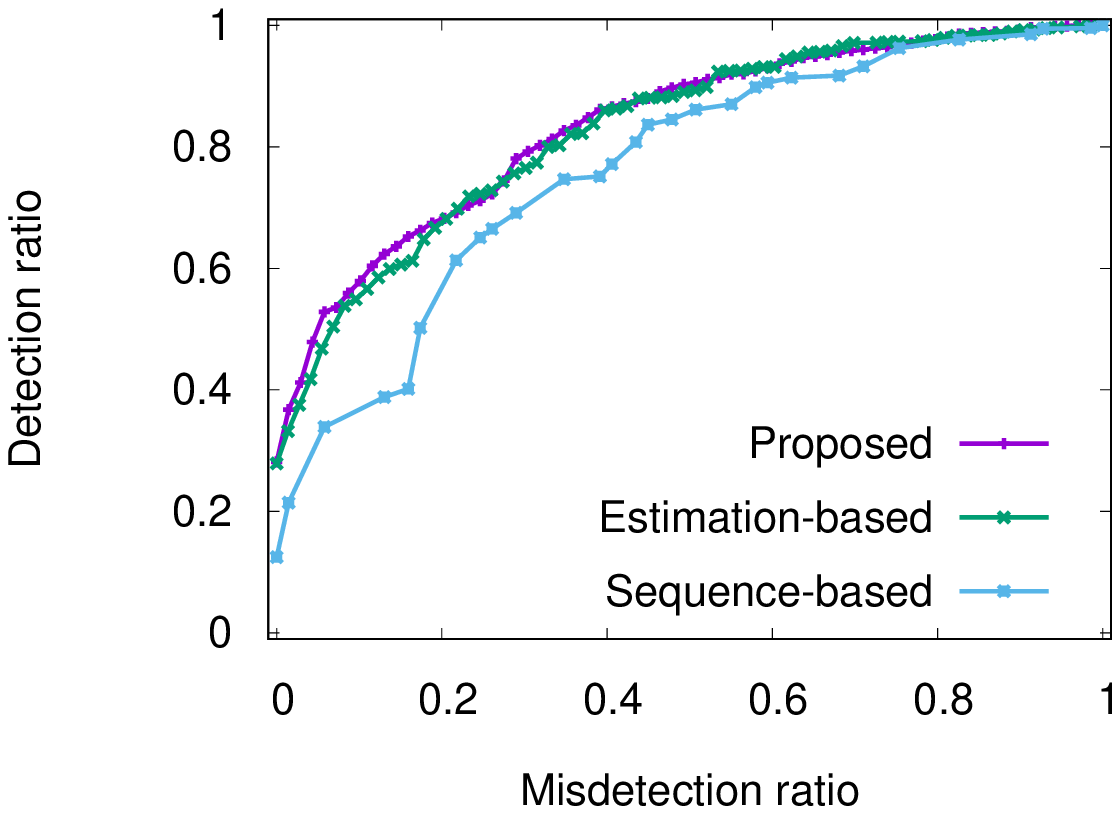}}
\\
	\subfloat[$A_5$]{\includegraphics[width=\subfigwidth]{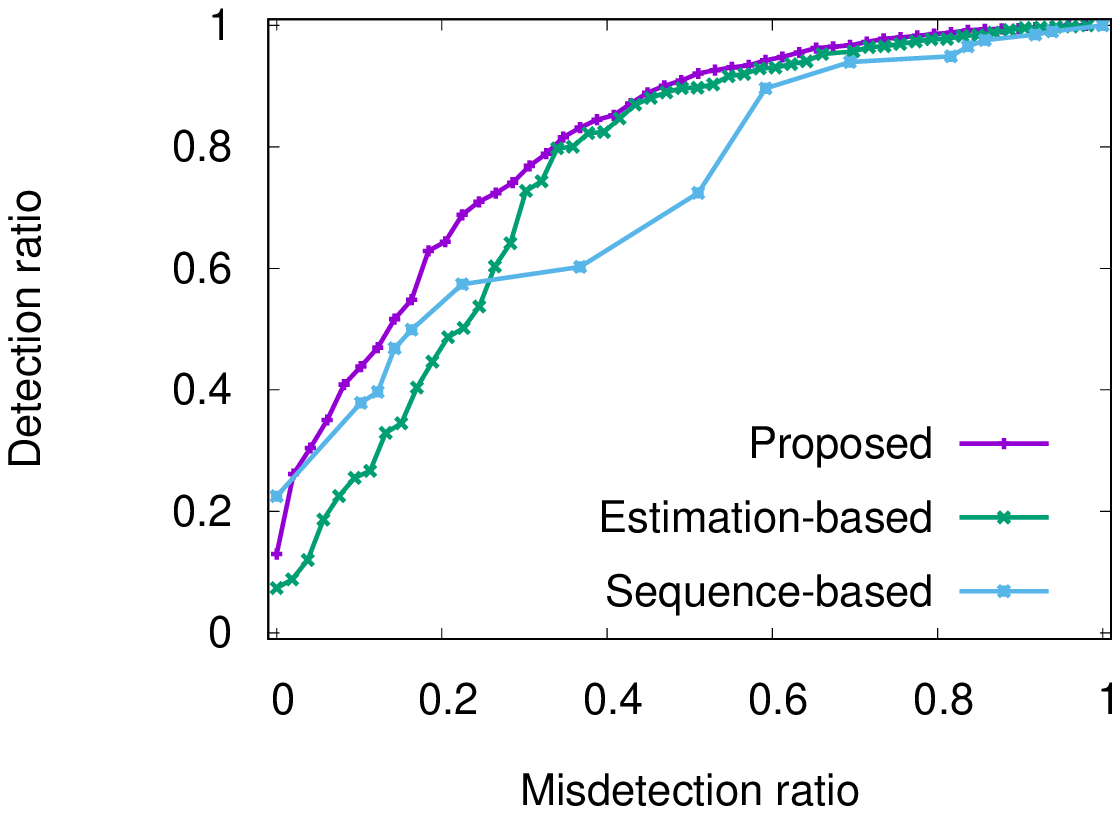}}
	\subfloat[$A_6$]{\includegraphics[width=\subfigwidth]{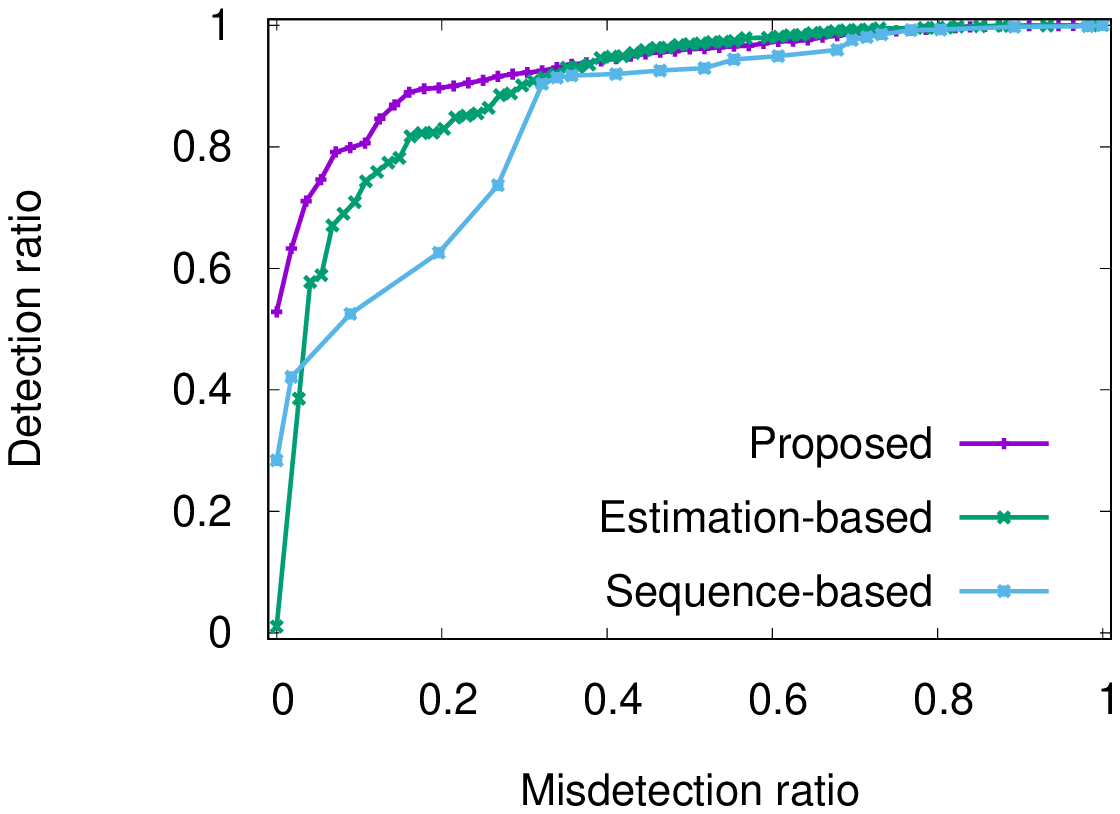}}
	\subfloat[$A_7$]{\includegraphics[width=\subfigwidth]{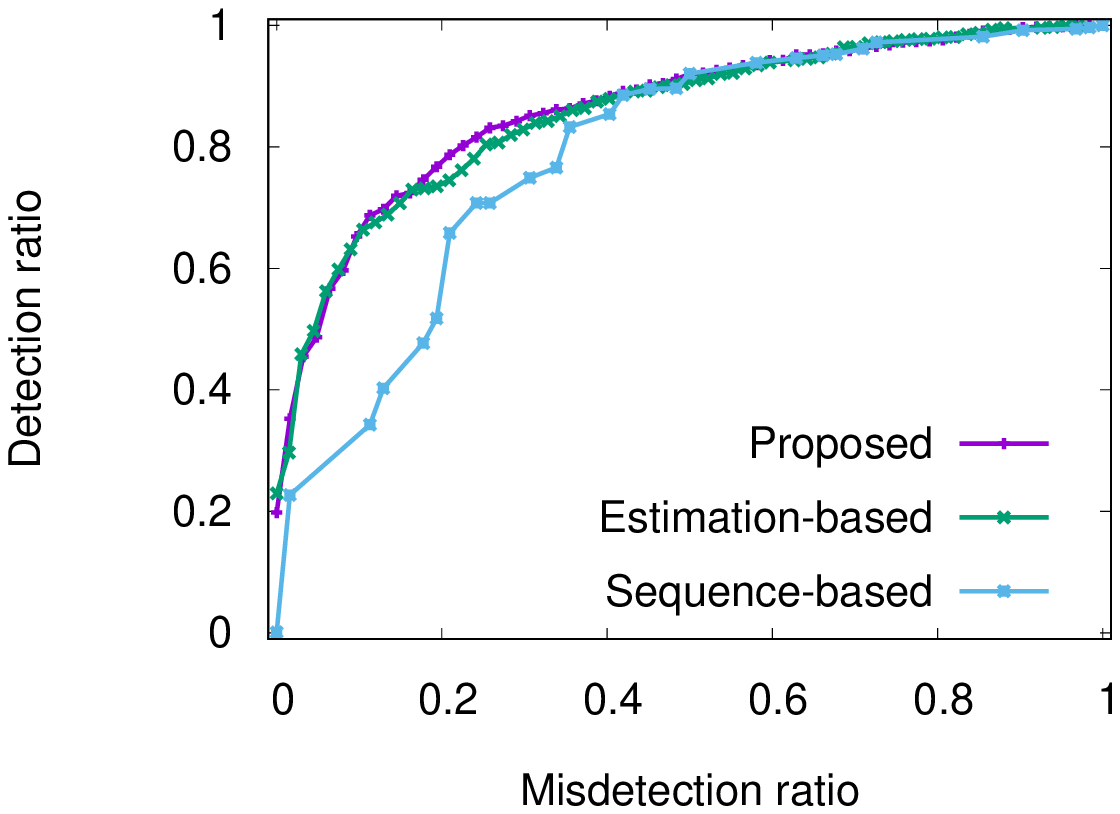}}
	\subfloat[$A_8$]{\includegraphics[width=\subfigwidth]{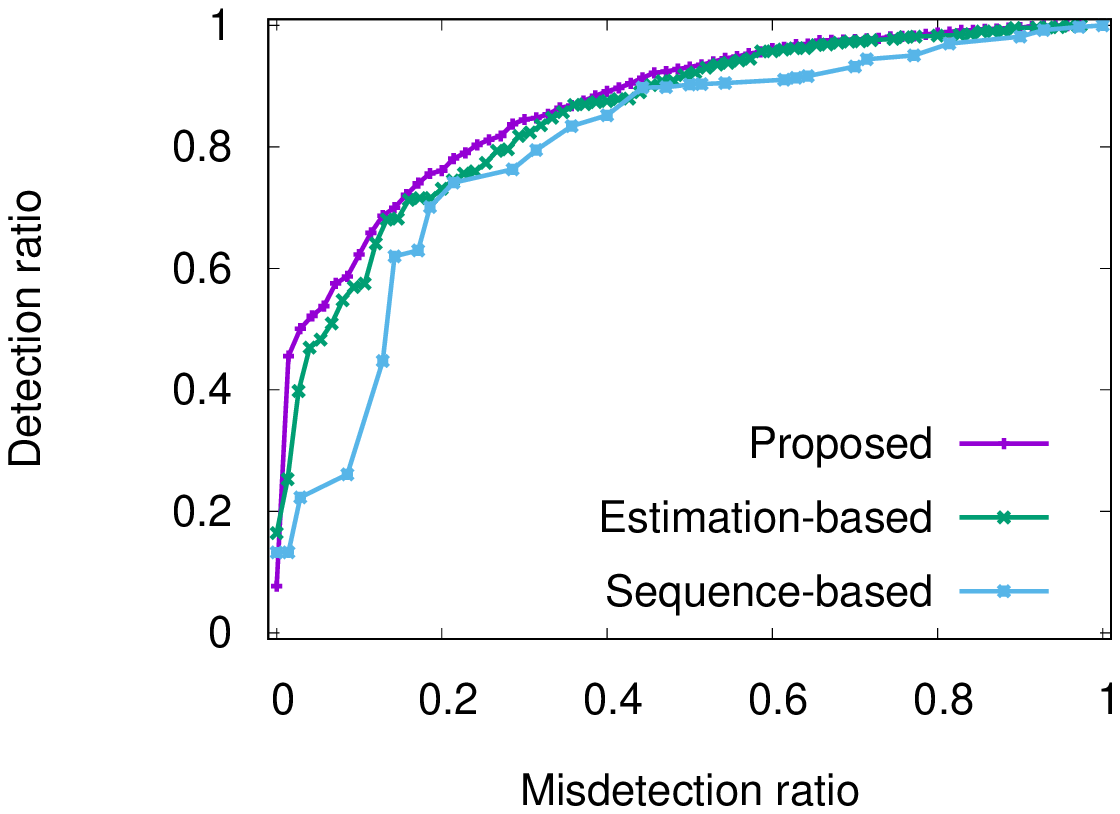}}
\\
	\subfloat[$A_9$]{\includegraphics[width=\subfigwidth]{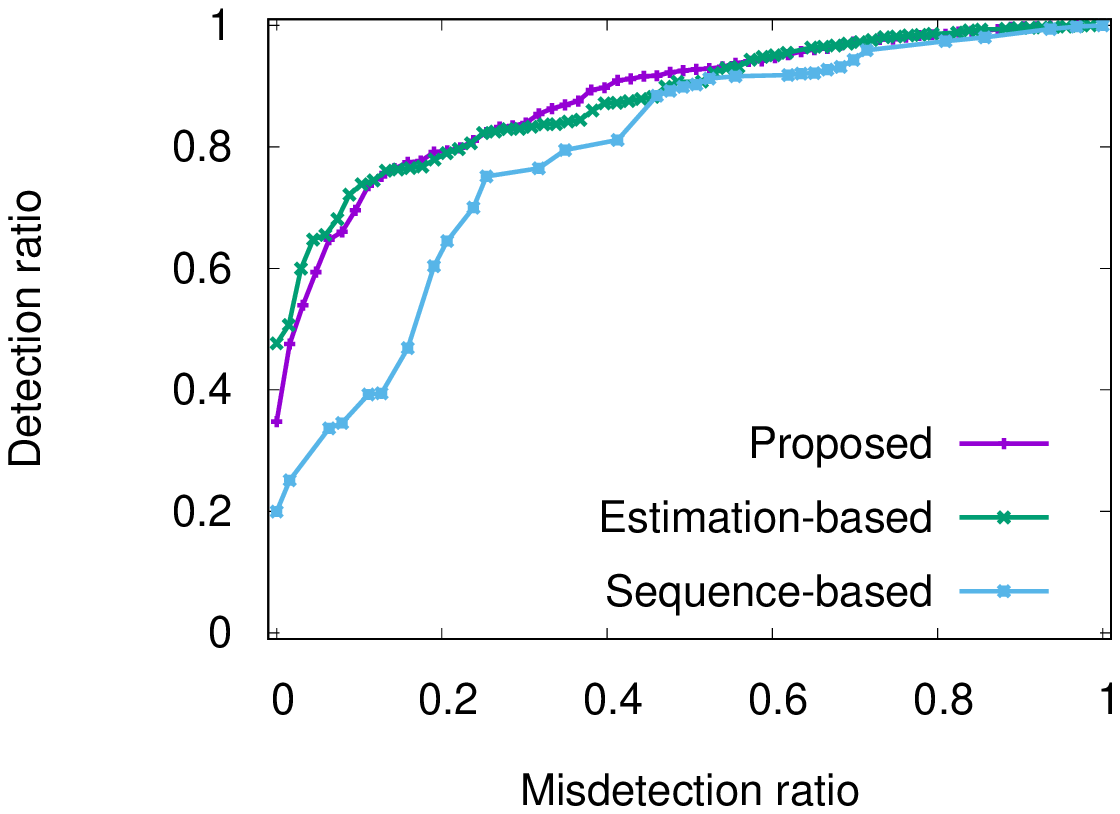}}
	\subfloat[$A_{10}$]{\includegraphics[width=\subfigwidth]{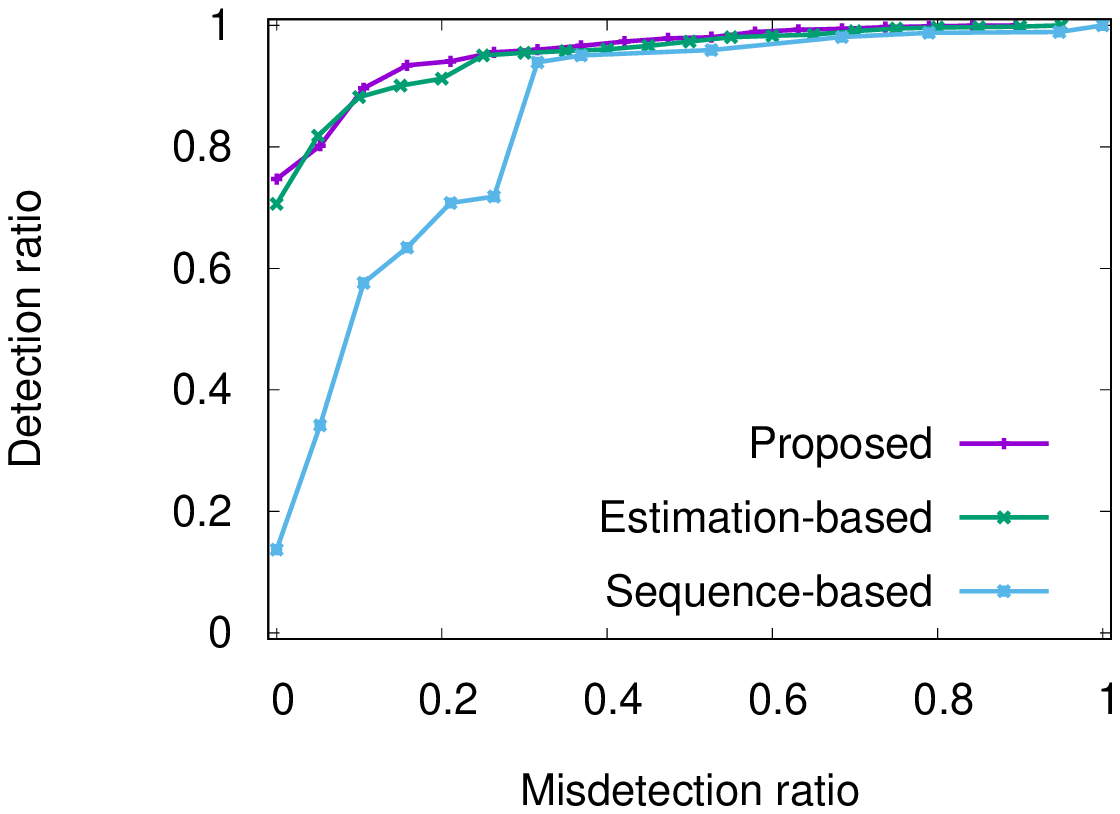}}
	\subfloat[$B_1$]{\includegraphics[width=\subfigwidth]{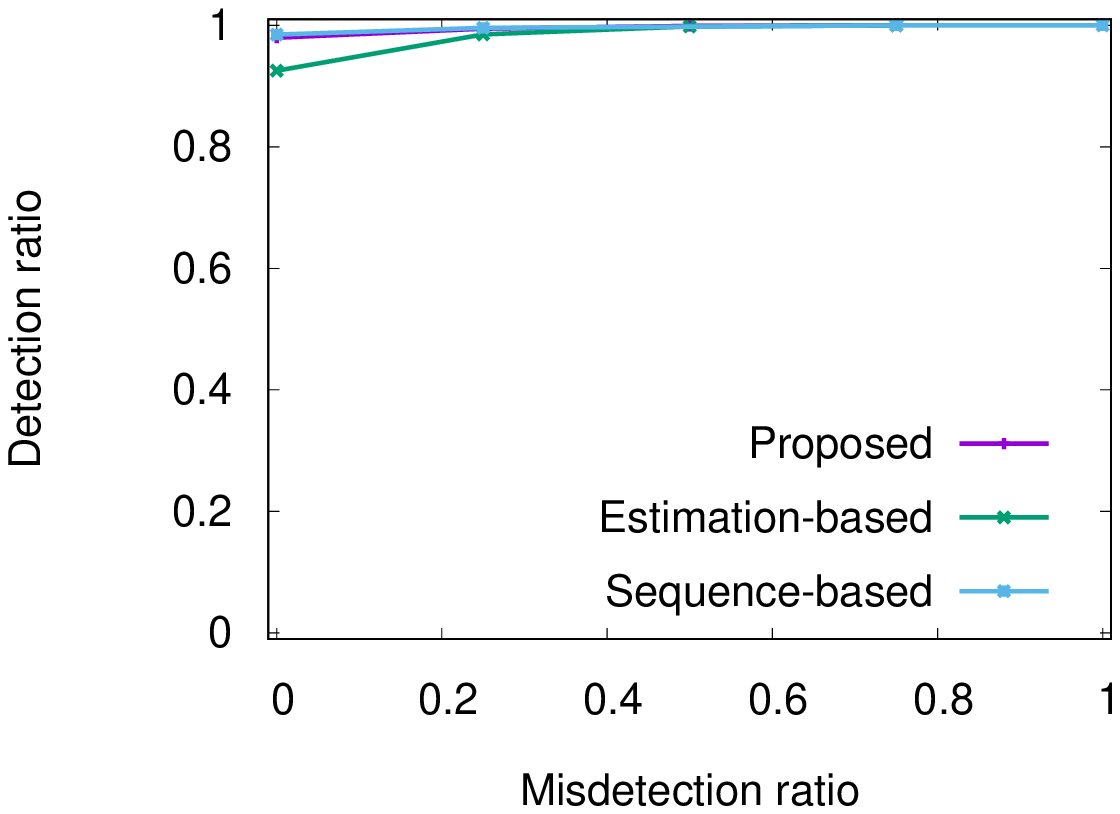}}
	\subfloat[$B_2$]{\includegraphics[width=\subfigwidth]{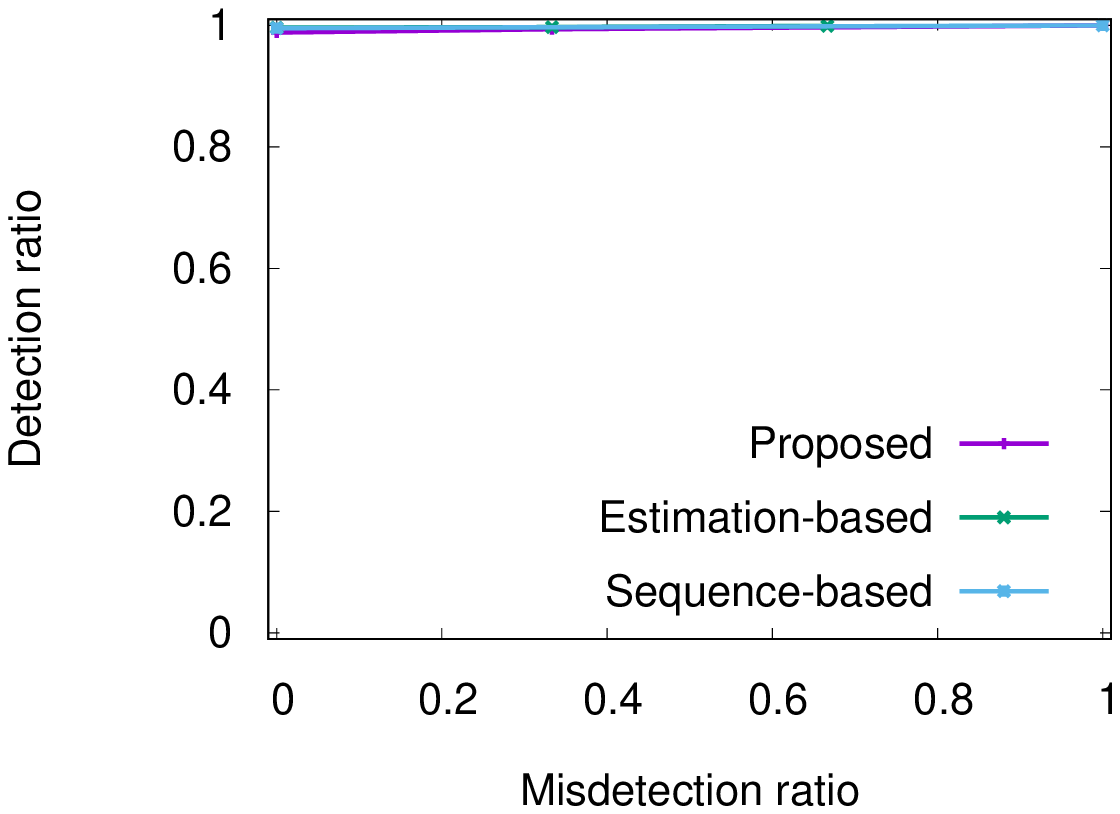}}
\\
	\subfloat[$B_3$]{\includegraphics[width=\subfigwidth]{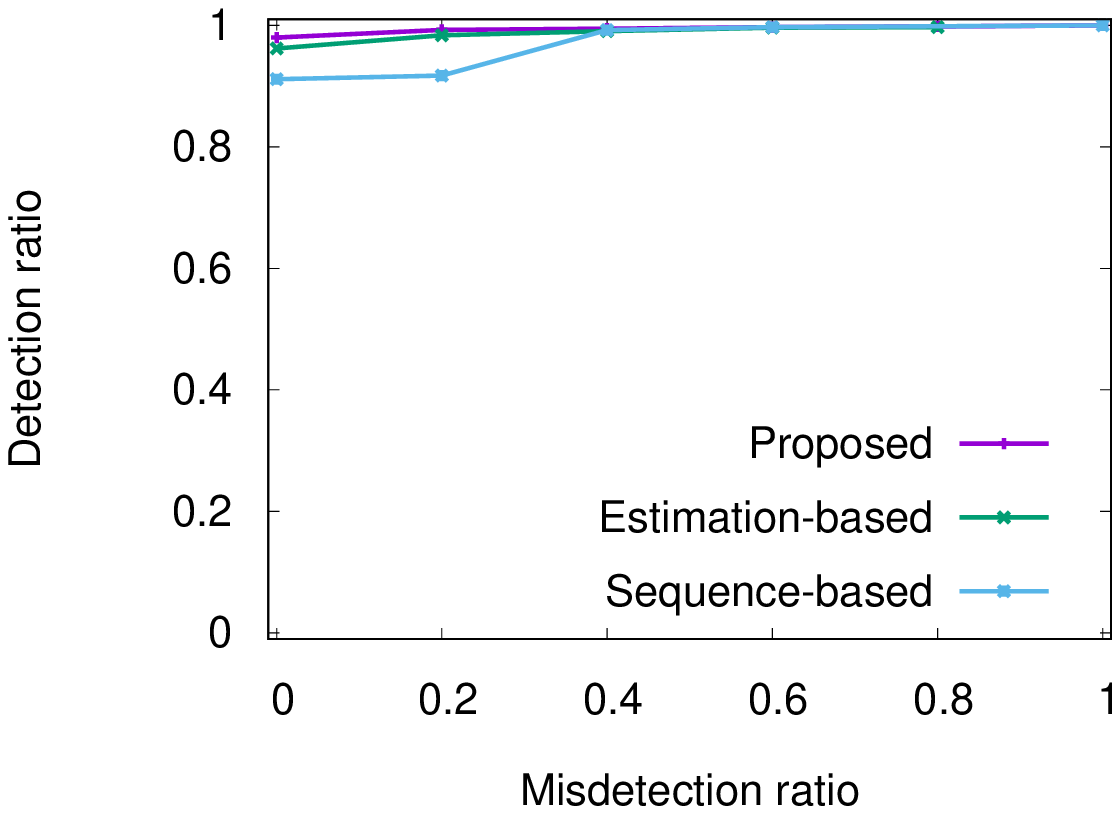}}
	\subfloat[$B_4$]{\includegraphics[width=\subfigwidth]{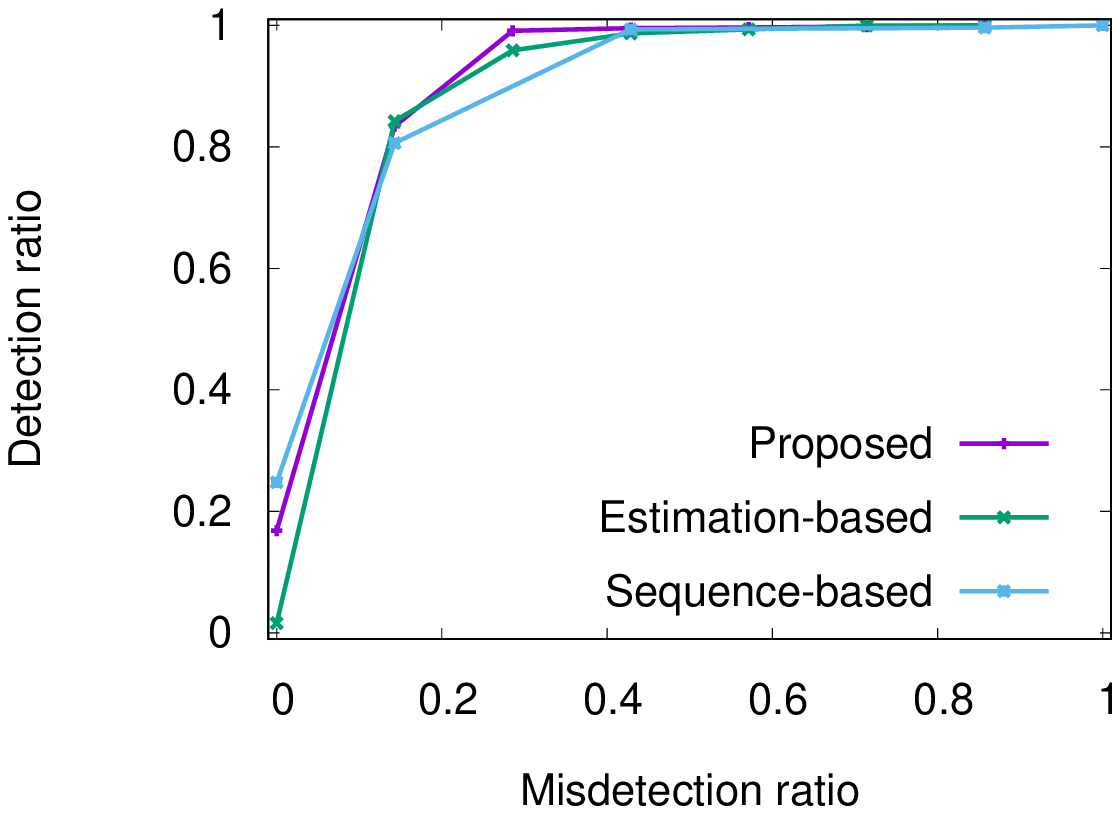}}
	\subfloat[$B_5$]{\includegraphics[width=\subfigwidth]{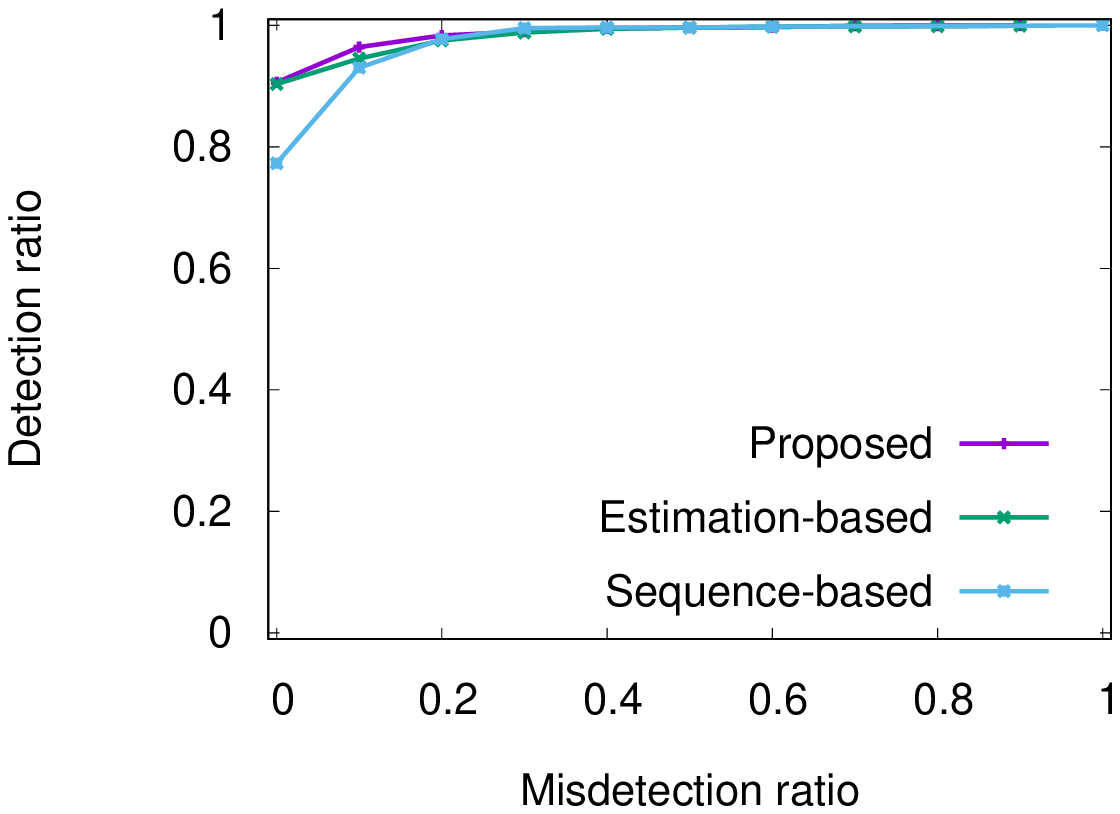}}
	\subfloat[$B_6$]{\includegraphics[width=\subfigwidth]{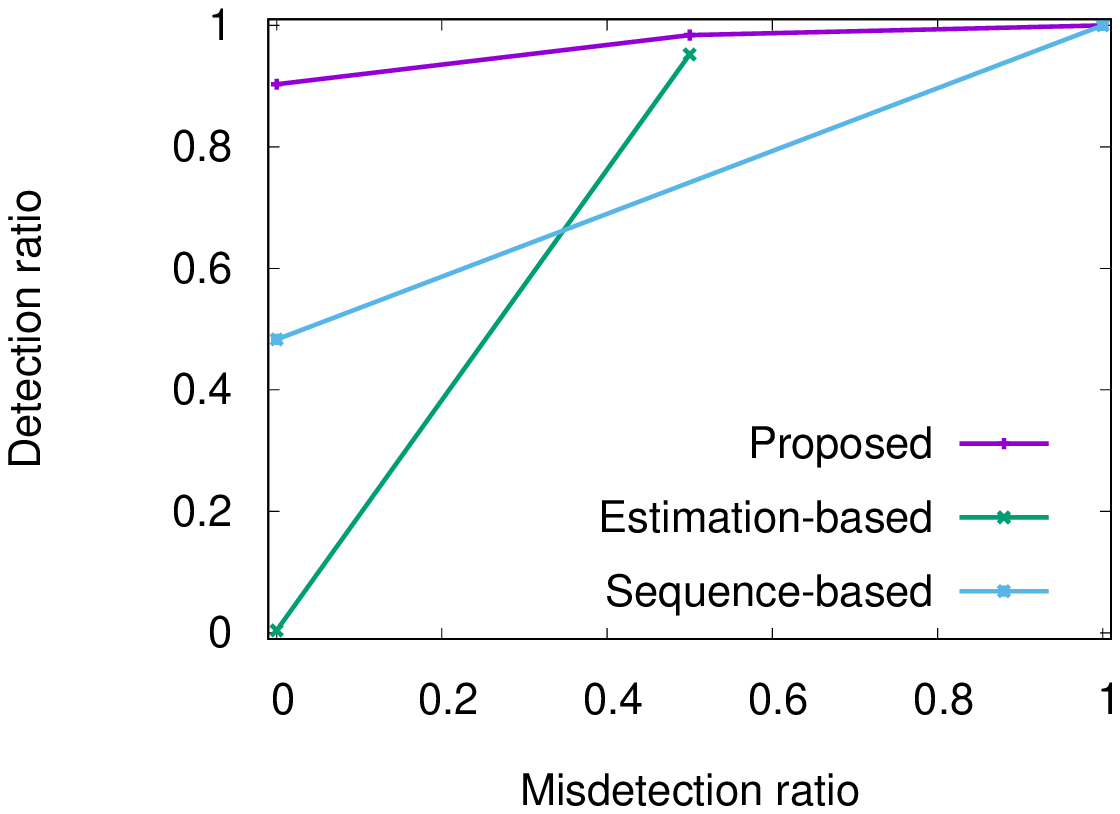}}
\\
	\subfloat[$B_7$]{\includegraphics[width=\subfigwidth]{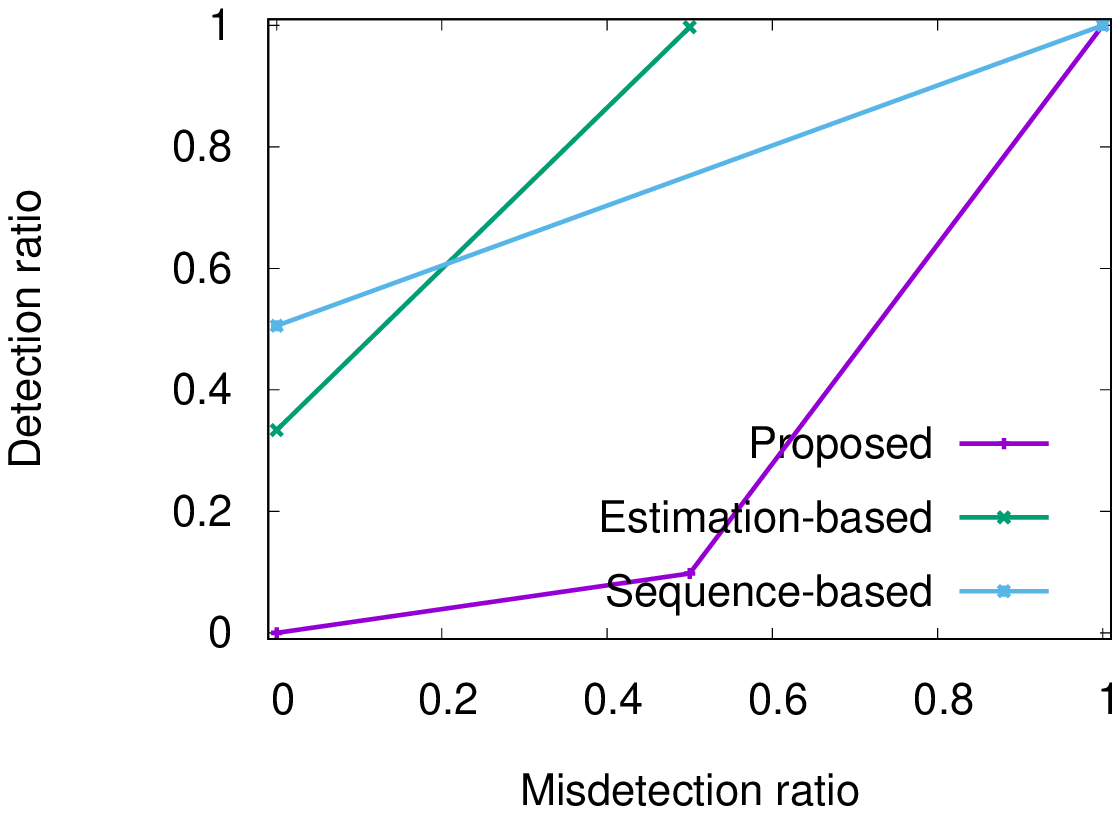}}
	\subfloat[$B_8$]{\includegraphics[width=\subfigwidth]{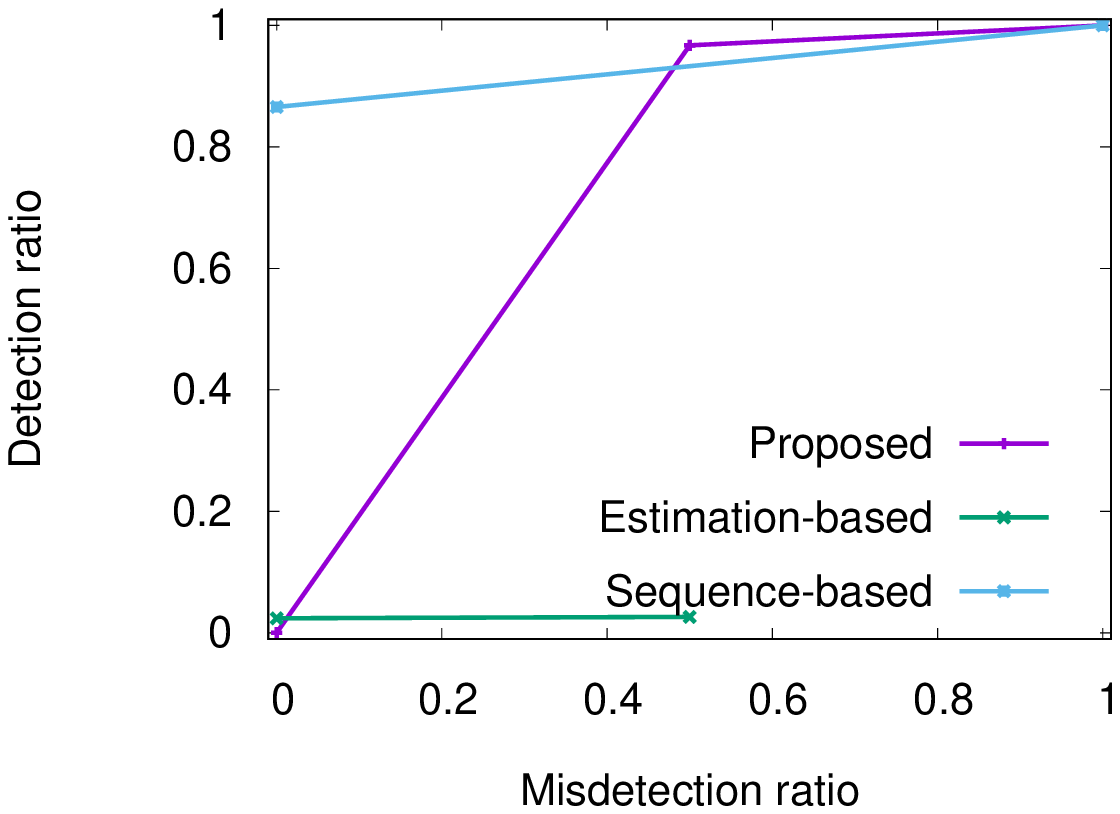}}
	\subfloat[$B_9$]{\includegraphics[width=\subfigwidth]{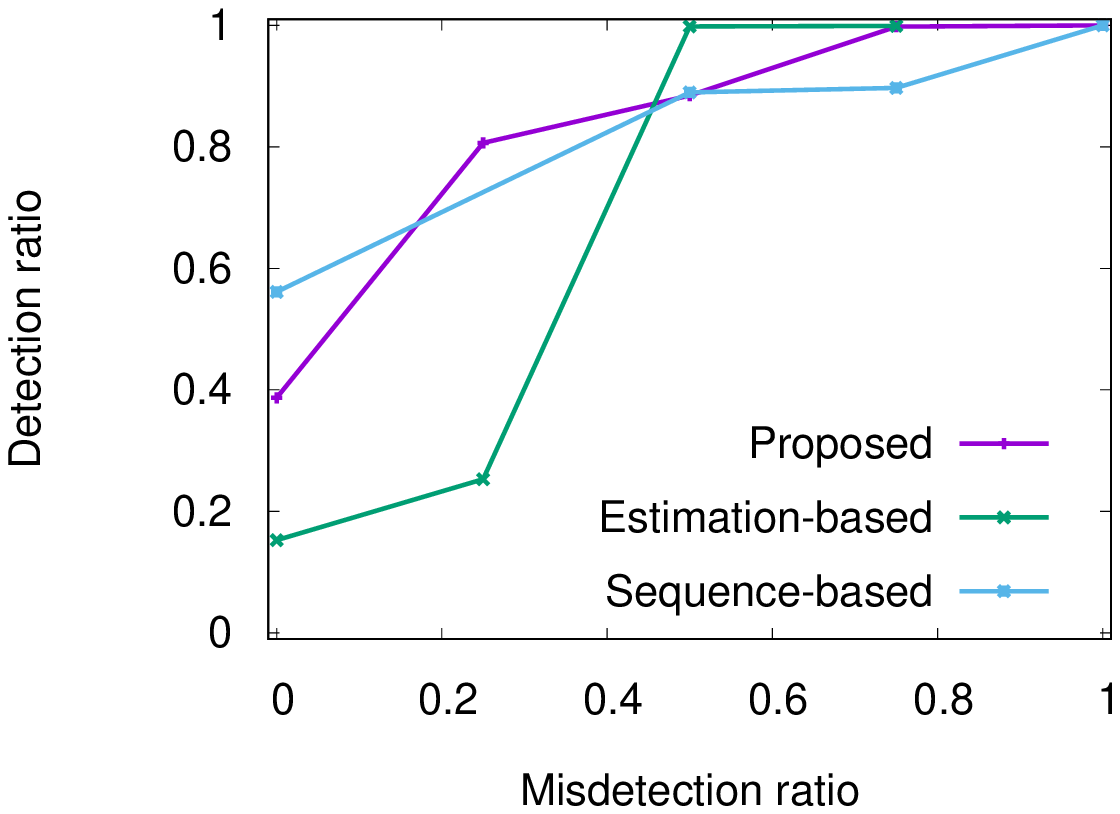}}
	\subfloat[$B_{10}$]{\includegraphics[width=\subfigwidth]{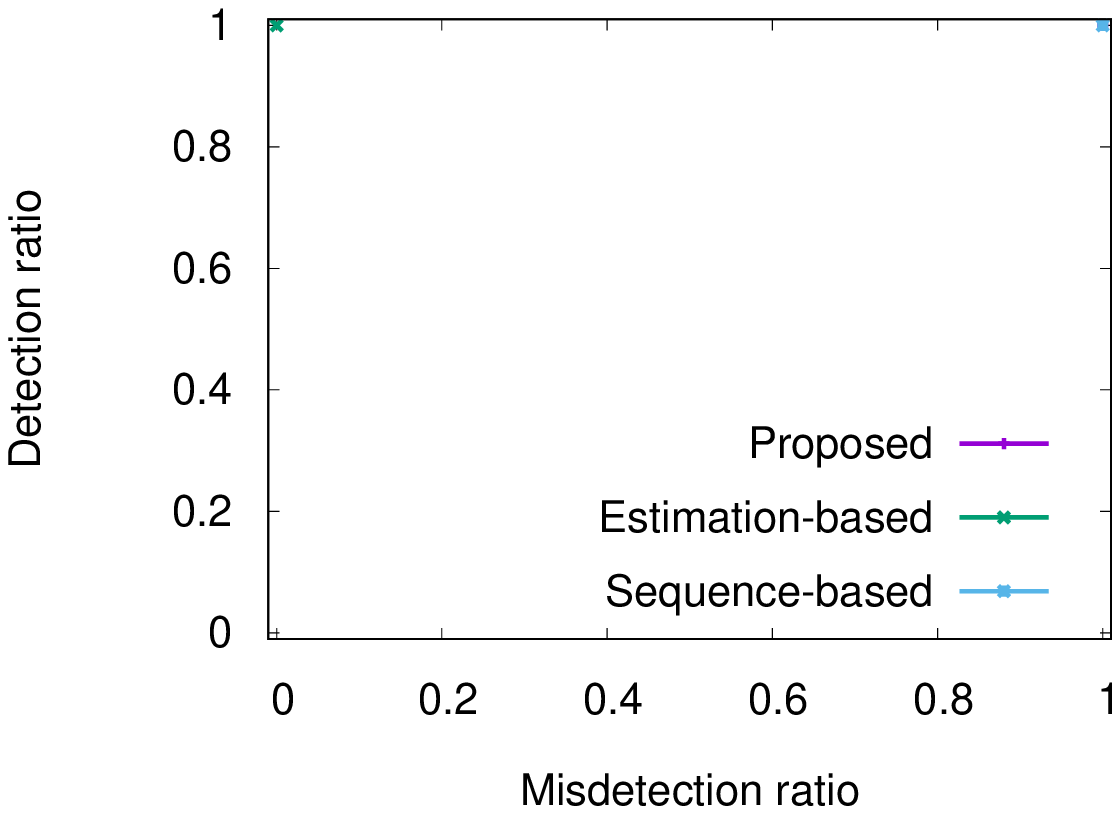}}
	\caption{Detection results for each month's data of each home.}
	\label{fig:roc_each}
\end{figure*}

The evaluation results of the proposed and compared methods for each month are shown in Fig.~\ref{fig:roc_each}.

The proposed method achieved a higher detection ratio with the same misdetection ratio of the sequence-based method in the case of home~$A_1$, $A_2$,~$\dots$,~$A_{10}$.
In particular, compared with the highest detection ratios having less than 10\% misdetections, the proposed method achieved a 46.0\% higher detection ratio than the sequence-based method in home~$A_{10}$.
This occurred because the proposed method has a narrower time range regarded as legitimate than the sequence-based method.
When the sequence-based method tries to reduce misdetections, the legitimate range is expanded as the learning data increases because the devices are operated at various times of day.
In contrast, because the proposed method estimates the state according to the state of each day, the legitimate time range can be narrowed.
Furthermore, the detection of single operations by the proposed method is another reason for its improved performance; the single operation means that there are no other operations before $T_{\textit{seq}}$~s.
When the sequence-based method cannot use short-term information, it cannot determine legitimate/anomalous operations because it must learn from only the time-of-day information.
Because the proposed method learns from the short- and long-term information, it can determine whether single operations are legitimate or anomalous from the long-term information.
However, the detection results of the methods were almost the same in homes~$B_1$, $B_2$, $B_3$, $B_4$, and~$B_5$.
In these cases, the operations were performed at the same time of day and were included in the same sequences.
Thus, the sequence-based method learned the behaviors accurately.

The proposed method also achieved a higher detection ratio with the same misdetection ratio of the estimation-based method in the case of homes~$A_1$, $A_2$, $A_3$, $A_5$, and~$A_6$.
In particular, compared with the highest detection ratios having less than 10\% misdetections, the proposed method achieved a 15.4\% higher detection ratio than did the estimation-based method in home~$A_3$.
This occurred because the proposed method can learn the relations between operations of the cooking stove and the operations of other frequently used devices, which included air conditioners, heaters, room lights, washing machines used in the morning, and refrigerators.
As an example of a legitimate behavior, a user might turn off a heater before using the cooking stove in order to regulate the ambient temperature.
The estimation-based method only determines whether the users are about to cook.
When users operated non-cooking devices, the probability of cooking was only slightly increased, and the estimation-based method could not determine the state.
However, the proposed method can learn the behavior sequence including the operations of such devices to grasp the legitimate operations of the cooking stoves.
In contrast, when there were fewer operations related to non-cooking devices, such as in homes~$A_4$, $A_7$, $A_8$, $A_9$, $A_{10}$, $B_1$, $B_2$, $B_3$, $B_4$, and~$B_5$, the detection results of the proposed method were slightly improved.
During the recorded months, devices such as heaters and air conditioners were not used, and their operations were almost always single-use or used with cooking equipment.
Therefore, nearly all operations of the cooking stoves could be determined as legitimate or not by estimating the home states.

The detection results of all methods were not stable in homes~$B_6$, $B_7$, $B_8$, $B_9$, and~$B_{10}$.
The numbers of operations included in these cases were too small to train the behavior models sufficiently.
However, the misdetections in those homes were not significant because there were only a small number of operations.


\section{Conclusion and Future Works}\label{sec:Conclusion}
To detect anomalous operation attacks on IoT devices in a home, we proposed a detection method that estimates the home state based on the observed values of IoT sensors and device operations and learns the event sequences of users in the home in each estimated state.
After training, when a device operation is observed to determine whether it is legitimate or anomalous, the proposed method calculates the occurrence probability of the sequence related to the target operation.
If the occurrence probability is lower than the threshold, the operation is detected as anomalous.
For this evaluation, we simulated anomaly detection using behavioral logs and sensor data obtained from real homes for one month.
We evaluated the improvements of the proposed method and the effectiveness of each part by comparing the proposed method to other methods, one of which did not use sequence information and the other did not estimate the in-home situation.
We found that the proposed method achieved a 15.4\% higher detection ratio with fewer than 10\% misdetections by using the sequence information, and it achieved a 46.0\% higher detection ratio with fewer than 10\% misdetections by using the estimation of the in-home situation.
Thus, the proposed approach can analyze the legitimate behavior of users and legitimate usages of the IoT devices comprehensively by using long- and short-term information, that is, by estimating the home state transition and using the sequence of behaviors.
However, a certain amount of data was required to learn the behaviors of users in the home.

In this study, we simulated the proposed method by setting a cooking stove as the target device.
Evaluating the proposed method when other devices are used as detection targets remains as a future task.
Furthermore, although we used data for one month for this evaluation, another future task will involve collecting data for a longer period of time and from many actual homes to verify the utility of the method.


\acknowledgement{
This work was supported by the Mitsubishi Electric Cybersecurity Research Alliance Laboratories.
They supported collection, analysis, and interpretation of data and the design study of estimating in-home situations.

This work was also supported by JSPS KAKENHI Grant Number JP21J12993.
It supported designing the method to combine the estimation of in-home situations and behavior sequences and the writing of this paper.

We would like to thank Editage (www.editage.com) for English language editing.
}
\conflictofinterest{Authors state no conflict of interest.}


\end{document}